\documentclass[fleqn,usenatbib,onecolumn]{mnras}

\usepackage{newtxtext,newtxmath}
\usepackage[T1]{fontenc}

\DeclareRobustCommand{\VAN}[3]{#2}
\let\VANthebibliography\thebibliography
\def\thebibliography{\DeclareRobustCommand{\VAN}[3]{##3}\VANthebibliography}

\usepackage{graphicx}	% Including figure files
\usepackage{amsmath}	% Advanced maths commands

\usepackage{amssymb}	% Extra maths symbols
\usepackage{subfigure}
\usepackage{parskip}
\usepackage{xcolor}

\newcommand{\bea}{\begin{eqnarray}}
\newcommand{\eea}{\end{eqnarray}}
\newcommand{\be}{\begin{equation}}
\newcommand{\ee}{\end{equation}}

\newcommand{\vc}[1]{\mbox{\boldmath $#1$}}

\title[TDSL Muztagh-Ata $1.93$m]{Forecast of observing time delay of the strongly lensed quasars with Muztagh-Ata $1.93$m telescope}

\author[S. Zhu et al.]{
Shanhao Zhu,$^{1}$
Yiping Shu$^{2,3}$,
Haibo Yuan,$^{1}$
Jian-Ning Fu,$^{1}$
Jian Gao,$^{1}$
Jianghua Wu,$^{1}$
\newauthor
Xiangtao He,$^{1}$
Kai Liao,$^{4}$
Guoliang Li$^{5}$,
Xinzhong Er,$^{6}$,
and Bin Hu$^{1}$\thanks{E-mail: bhu@bnu.edu.cn}
\\
$^{1}$Department of Astronomy, Beijing Normal University, Beijing 100875, China\\
$^{2}$Max-Planck-Institut f\"{u}r Astrophysik, Karl-Schwarzschild-Str. 1, 85748 Garching, Germany\\
$^{3}$Ruhr University Bochum, Faculty of Physics and Astronomy, Astronomical Institute (AIRUB),\\ German Centre for Cosmological Lensing, 44780 Bochum, Germany\\ 
$^{4}$School of Physics and Technology, Wuhan University, Wuhan 430072, China\\
$^{5}$Purple Mountain Observatory, Chinese Academy of Sciences, Nanjing, Jiangsu, 210023, China\\
$^{6}$South-Western Institute for Astronomy Research, Yunnan University, Kunming, 650500, China\\
}

\date{Accepted XXX. Received YYY; in original form ZZZ}

\pubyear{2021}

\begin{document}
\label{firstpage}
\pagerange{\pageref{firstpage}--\pageref{lastpage}}
\maketitle

\begin{abstract}
As a completely independent method, the measurement of time delay of strongly lensed quasars (TDSL) are crucial to resolve the Hubble tension. Extensive monitoring is required but so far limited to a small sample of strongly lensed quasars. 
Together with several partner institutes, Beijing Normal University is constructing a 1.93m reflector telescope at the Muztagh-Ata site in west China, which has the world class observing conditions with median seeing of 0.82 arcsec and median sky brightness of 21.74 $\mathrm{mag} \operatorname{arcsec}^{-2}$ in V-band during the dark time. 
The telescope will be equipped with both a three-channel imager/photometer which covers $3500-11000$ Angstrom wavelength band, and a low-medium resolution ($\lambda/\delta\lambda=500/2000/7500$) spectrograph.  
In this paper, we investigate the capability of Muztagh-Ata 1.93m telescope in measuring time delays of strongly lensed quasars. We generate mock strongly lensed quasar systems and light curves with microlensing effects based on five known strongly lensed quasars, {\it i.e.}, RX J1131-1231, HE 0435-1223, PG 1115+080, WFI 2033-4723 and SDSS 1206+4332. In particular, RX J1131-1231 is generated with lens modeling in this work. Due to lack of enough information, we simulate the other 4 systems with the public data without lens modeling. 
According to simulations, for RX J1131-like systems (wide variation in time delay between images) the TDSL measurement can be achieved with the precision about $\Delta t=0.5$ day with 4 seasons campaign length and 1 day cadence. This accuracy is comparable to the up-coming TDCOSMO project. And it would be better when the campaign length keeps longer and with high cadence. As a result, the capability of Muztagh-Ata 1.93m telescope allows it to join the network of TDSL observatories.
It will enrich the database for strongly lensed quasar observations and make more precise measurements of time delays, especially considering the unique coordinate of the site. 
\end{abstract}


%%%%%%%%%%%%%%%%%%%%%%%%%%%%%%%%%%%%%%%%%%%%%%%%%%

%%%%%%%%%%%%%%%%% BODY OF PAPER %%%%%%%%%%%%%%%%%%

\section{Introduction}

The Hubble constant ($H_0$) is an important parameter for characterizing the current expansion rate of the universe. However, there is a serious discrepancy on the measured Hubble constant value between different methods~\cite{Freedman:2017yms,Aghanim:2018eyx,Abbott:2017smn,Riess:2019cxk,2019ApJ...882...34F}. 
One of the most traditional methods of determining $H_{0}$, namely the distance ladder, uses three different distance indicators ranged from nearby milky way to the faraway cosmological scales. This method utilize the parallax measurement, Cepheid variables~\cite{Riess:2019cxk}, the tip of red-giant branch stars (TRGB)~\cite{2019ApJ...882...34F} and Type Ia supernovae. The most recent controversial measurements gives the best-fit value of $H_0=73.2 \pm 1.3 \mathrm{~km} \mathrm{~s}^{-1} \mathrm{Mpc}^{-1}$~\cite{H0_ladder}, a $4.2\sigma$ in tension with the Planck cosmic microwave background (CMB) observations under $\Lambda$CDM cosmology, in which $H_0=67.4 \pm 0.5 \mathrm{~km} \mathrm{~s}^{-1} \mathrm{Mpc}^{-1}$~\cite{H0_Planck}.
The CMB and baryon acoustic oscillation (BAO) methods currently yield lower values of $H_{0}$, while Cepheids yield the highest values and TRGB results falling in the middle \cite{H0_tension}. 
In order to figure out whether the tension is due to unaccounted systematic errors, or the existence of ``new physics'', we need independent measurements with accuracy better than $2\%$~\cite{Verde:2019ivm}. One of the promising approach is to use the time delay between multiple images of strong lensing~\cite{TD_good1,TD_good2,H0li_13}, namely ``time-delay cosmography''. 
Light from a distant object is splitted and produces multiple images, when it intervenes massive objects along its path. 
As light travels in different paths and feels different gravitational forces, the light of images do not always reach the observer at the same time and it causes time delays. 
The time delays among images are affected not only by the mass distribution in the lens plane and the projected mass along the line of sight, but also depends on the geometry of lens system. Hence, an accurate measurement of the time delay helps to measure the Hubble constant.

In 1964, Refsdal proposed the ideal of using gravitational lensing time delay as a tool to measure $H_{0}$ \cite{Refsdal}. The first actual strong lensing measurement is done by \citet{first_SL}. 
It was a quasar lensing system (Q0957+561) at redshift $z=1.4$. 
The first measurement of time delay was made by \cite{first_td}, and was later confirmed by \cite{confirm_td}.
Quasars are ideal sources for ``time-delay cosmography'' thanks to their high luminosity and variability. 
To date, more and more strongly lensed quasar systems have been discovered in various surveys~\cite{2006AJ....132..999O,2012AJ....143..119I,2016MNRAS.456.1595M}.
More importantly, some of the lensed quasar systems have been used to measure $H_0$~\cite{H0Li_9,H0li_13}. There are several teams focus on this topic, such as COSMOGRAIL\footnote{cosmograil.org}, H0LiCOW\footnote{h0licow.org} and STRIDES\footnote{strides.astro.ucla.edu}. 
The COSMOGRAIL project began in 2004 with a mission to monitor strongly lensed quasars and measure the time delays. 
The collaboration monitored dozens of lensed quasars with six 1-2m class telescopes all around the world. 
This network is constituted by the Swiss 1.2 m Euler telescope located at La Silla, Chile; the Swiss-Belgian 1.2 m Mercator telescope, located in the Canaria islands (La Palma, Spain); the 2 m robotic telescope of the Liverpool University (UK) at La Palma; the 1.5 m telescope of Maidanak observatory in Uzbekistan; the 2 m Himalayan Chandra Telescope (HCT) in Hanle, Indian ; and 2m2 MPG/ESO telescope at La Silla.

%%%%%%%%%%%%%%%%%%%%%%%%%%%%%%%%%%%%%%%%%%
\begin{figure}
 \includegraphics[width=0.8\columnwidth]{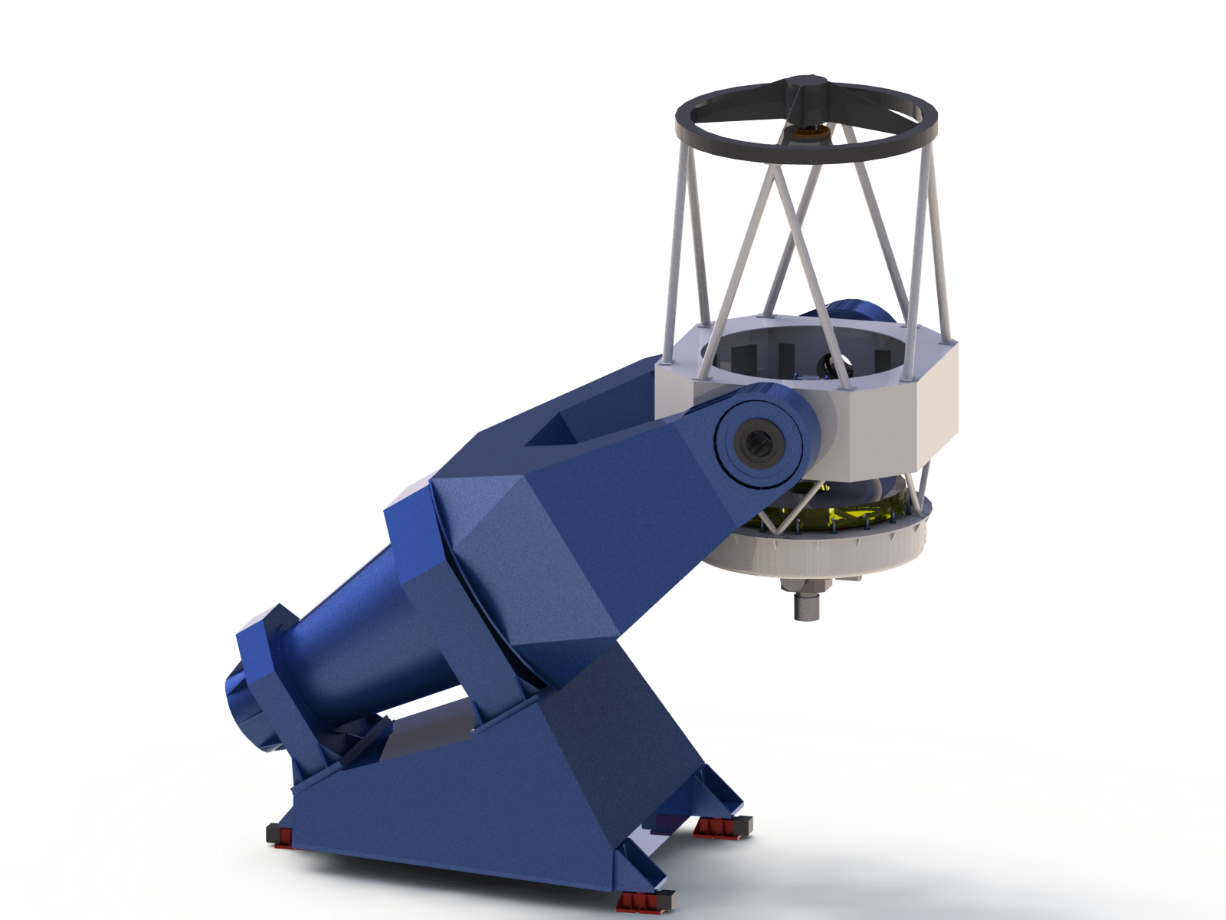}
 \caption{Conceptual design of Muztagh-Ata $1.93$m telescope.}
 \label{fig:telescope}
\end{figure}

%%%%%%%%%%%%%%%%%%%%%%%%%%%%%%%%%%%%%%%%%%
\begin{figure}
    \includegraphics[width=\columnwidth]{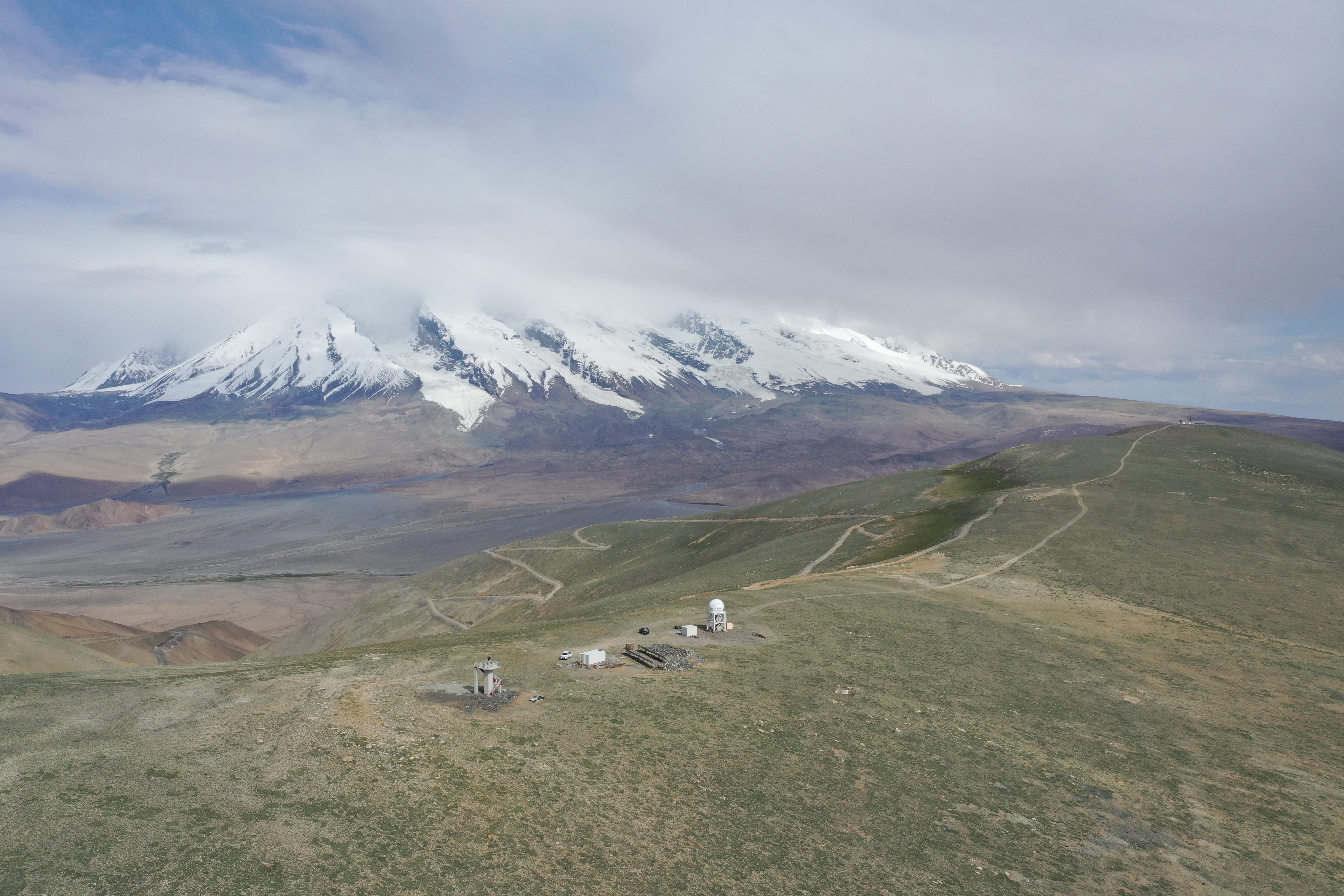}
    \caption{Full view of the site (altitude: 4526m). The white dome in the middel is a 50cm telescope investigated by Beijing Normal University. 
    The 1.93m telescope is planned to be mounted next to the 50cm one.
    The background mountain is Muztagh-Ata (altitude: 7509m).}
    \label{fig:site}
\end{figure}

A 1.93m reflector telescope equipped with both a three-channel imager/photometer and a low-medium resolution spectrograph is currently under the construction at Muztagh-Ata site and will be finished in 2-3 years.
And Figure \ref{fig:telescope} is the conceptual design of it.
The telescope is mainly invested by Beijing Normal University (BNU) and cooperated with Xinjiang Astronomical Observatory (XAO), Nanjing Institute of Astronomical Optics and Technology of Chinese Academy of Sciences (NIAOT) and Xinjiang University (XJU).
The photometry wavelength band covers $3500-11000$ Angstrom and spectrograph has three resolution, $\delta\lambda/\lambda=500/2000/7500$. The field of view is $20$ arcmin with help of the correction mirror. The 300s exposure $10\sigma$ limiting magnitude in V-band is $23.79$. 
The telescope is designed in the R-C optical system with three focus, namely the Cassegrain focus, the declination axis focus and the coud\'{e} focus. 
The effective aperture of the telescope is 1.93 meter and the focal ratio is $f/8$. The pixel size of CCD is 13.5 $\mu m$. The scale on the focal plane is $0.183''/{\rm pix}$ and the quantum efficiency is about 0.95. The guiding system can keep the tracking precision at the 0.3 arcsec level within 2 hours. The pointing precision is expected to be 5 arcsec with the pointing model correction. And it can be improved to the 1 arcsec level after the secondary correction.

Muztagh-Ata site is located at $38^\circ 19' 47''$N and $74^\circ 53' 48''$E in the southwest of Xinjiang Uygur Autonomous Region of China, with an altitude of 4526 meters. The full view of the site is presented in Figure~\ref{fig:site}. It is one of the best astronomical sites in China. 
The seeing median value is 0.82 arcsec \cite{2020RAA....20...87X}. 
The median value of the sky brightness is 21.35 $\mathrm{mag} \operatorname{arcsec}^{-2}$ in V-band during the nighttime. For the case without moon, this number can be upgraded into 21.74 $\mathrm{mag}\operatorname{arcsec}^{-2}$ (V-band). The median of relative humidity is $49\%$ for nighttime and $39\%$ for daytime. The median value of nighttime wind speed is $5.5 \mathrm{ms}^{-1}$ and it is $6.5 \mathrm{ms}^{-1}$ for daytime \cite{2020RAA....20...86X}.
All these conditions makes the telescope ideal for time-domain astronomical researches.

In this paper, we forecast the capability of observing time delay of strongly lensed quasars (TDSL) with Muztagh-Ata $1.93$m telescope. The rest of the paper are structured as follows. In Section~\ref{sec:basics}, we introduce the lens modeling. Section~\ref{sec:simulation} describes the simulation process. The method of measuring time delays are given in Section~\ref{sec:measure}. In Section~\ref{sec:conclusion}, we arrive our conclusions. 
In this study, we adopt a flat $\Lambda$CDM cosmology model with parameter $\Omega_{m}=0.3$ and $h=0.7$.

%%%%%%%%%%%%%%%%%%%%%%%%%%%%%%%%%%%%%%%%%%
\section{Lens modeling}
\label{sec:basics} % used for referring to this section from elsewhere

In this section, we present the lens modeling, including lens basics, the lens mass distribution and brightness distribution.  

%%%%%%%%%%%%%%%%%%%%%%%%%%%%%%%%%%%%%%%%%%
\subsection{Lens basics}

We denote the angular diameter distances between the source and the lens as $D_{\mathrm{ds}}$, between the source and the observer as $D_{\mathrm{s}}$, and between the lens and the observer as $D_{\mathrm{d}}$. We introduce the angular coordinates in the image plane as $\vc\theta$, which are perpendicular to the line of sight, and angular coordinate in the source plane as $\vc\beta$. The coordinates in the image and source planes are related through the lens equation
\begin{equation}
\vc\beta=\vc\theta - \vc\alpha(\theta)
=\vc\theta-\nabla_\theta \psi(\theta),
\end{equation}
where $\vc\alpha(\theta)$ is the deflection angle, $\psi(\theta)$ is the effective lens potential and $\nabla_\theta$ is the gradient in the image plane with respect to $\theta$. The lens potential is determined by the dimensionless projected surface mass density $\kappa$, also the lensing convergence
\begin{align}
\psi(\theta)&= \frac{1}{\pi} \int d^2\theta' \kappa(\theta') {\rm ln}|\theta-\theta'|;\\
\kappa(\theta)&=\Sigma(\theta)/\Sigma_{\rm cr}, \qquad {\rm where}\qquad 
\Sigma_{\rm cr}=\dfrac{c^2}{4\pi G}\dfrac{D_{\mathrm{s}}}{D_{\mathrm{d}} D_{\mathrm{ds}}}
\end{align}
is the critical surface mass density depending on the angular diameter distances. $\Sigma(\theta)$ is surface mass density of the lens. To the lowest order, the image distortions caused by gravitational lensing are described by the shear $\gamma$. The magnification for a point source is given by 
\begin{equation}
    \mu= \frac{1}{(1-\kappa)^{2}-\gamma^{2}}.
    \label{eq:mu}
\end{equation}

If the magnification $\mu$ is finite, only one image can be produced from a source. To produce multiple images, the source must cross an infinite magnification curve, which corresponds to a denominator of 0 in Eq. (\ref{eq:mu}) and divides the region where the new image is generated. Such curves are called critical curves in the image plane and caustics in the source plane.
The arrival time between multiple images generated by strong lensing is 
\be
\Delta t=\frac{1+z_{d}}{c} \frac{D_{\mathrm{d}} D_{\mathrm{s}}}{D_{\mathrm{ds}}}\left[\tau\left(\boldsymbol{\theta}^{(1)} ; \boldsymbol{\beta}\right)-\tau\left(\boldsymbol{\theta}^{(2)} ; \boldsymbol{\beta}\right)\right],
\label{eq:delta_t}
\ee
where $\tau(\boldsymbol{\theta} ; \boldsymbol{\beta})$ is the Fermat potential, and can be written as
\begin{equation}
    \tau(\boldsymbol{\theta} ; \boldsymbol{\beta})=\frac{1}{2}(\boldsymbol{\theta}-\boldsymbol{\beta})^{2}-\psi(\boldsymbol{\theta}).
    \label{eq:Fermat}
\end{equation}

When the source and the lens are perfectly aligned, the source is mapped to a ring image (the so called Einstein ring) and to a central image.  Einstein radius $\theta_E$ is the radius of Einstein ring, which reads
\begin{equation}
    \theta_{\mathrm{E}}=\sqrt{\frac{4GM\left(\theta_{\mathrm{E}}\right)}{c^{2}} \frac{D_{\mathrm{ds}}}{D_{\mathrm{d}} D_{\mathrm{s}}}},
    \label{eq:theta_E}
\end{equation}
where $M(\theta_{\mathrm{E}})$ is the mass within the Einstein radius. More details of the lens basics can be found in \cite{2006Gravitational}.

\subsection{Mass distribution and brightness distribution}

Power-law profile can provide a fairly good descriptions to the mass distribution of the realistic lens galaxies \cite{PEMD_good1,PEMD_good2}. The dimensionless surface mass density can be written as
\begin{equation}
    \kappa\left(\theta_{1}, \theta_{2}\right)=\frac{3-\gamma^{\prime}}{2}\left(\frac{\theta_{\mathrm{E}}}{\sqrt{q \theta_{1}^{2}+\theta_{2}^{2} / q}}\right)^{\gamma^{\prime}-1},
\end{equation}
where $\gamma^{\prime}$ is the radial power-law slope, $\theta_{\mathrm{E}}$ is the Einstein radius, and $q$ is the axis ratio of the ellipitical isodensity contours.

In addition to the lens galaxies, the external shear on the lens plane generally cannot be ignored in modeling. It can be written in the form of polar coordinates $r=\sqrt{\theta_1^2+\theta_2^2}$ and $\phi$ \cite{RXJ_paras}
\begin{equation}
    \psi_{\mathrm{ext}}(r, \varphi)=\frac{1}{2} \gamma_{\mathrm{ext}} r^2 \cos 2\left(\varphi-\phi_{\mathrm{ext}}\right),
\end{equation}
where $\gamma_{\mathrm{ext}}$ is the shear strength and $\phi_{\mathrm{ext}}$ is the shear angle.

S\'{e}rsic brightness profile is an empirical model verified by a large number of observations.
It has become the standard model for describing the surface brightness profiles of early-type galaxies and bulges of spiral galaxies \cite{Sersic_standard}. 
The brightness reads
\begin{equation}
    I(R)=I_{e} \exp \left\{-b_{n}\left[\left(\frac{R}{R_{e}}\right)^{1 / n}-1\right]\right\},
    \label{eq:sersic}
\end{equation}
where $n$ is the S\'{e}rsic index.
%which describes the shape of the light-profile. The smaller the index is, the less profile centrally concentrate. 
The parameter $b_{n}$ is a dimensionless parameter of about $2n-1/3$. $R_{e}$ is the half-light radius (also called effective radius) which means the luminosity within $R_{e}$ is half of the total stellar luminosity of the galaxy. And $I_{e}$ is the intensity at $R_{e}$, which can be calculated according to Eq.~(\ref{eq:sersic}) and the definition of $R_{e}$. The S\'{e}rsic index of most galaxies is between $1/2$ and 10. 
For elliptical galaxies, generally we have $n=4$, namely the de Vaucouleurs brightness distribution model \cite{De_Vau}. 

%%%%%%%%%%%%%%%%%%%%%%%%%%%%%%%%%%%%%%%%%%
\section{Simulation Process}
\label{sec:simulation}

In this section, we introduce the method for generating the mock light-curves of the strongly lensed quasars. 
The H0LiCOW collaboration \cite{H0li_13} used six strongly lensed quasar systems to constrain $H_0$. 
In this paper, we pick RX J1131-1231 as a working example to demonstrate the capability of Muztagh-Ata 1.93m telescope of measuring time delays. This is a system with a large time delay difference between images, but there has been a lot of modeling before.  
We use \texttt{lenstronomy} \cite{2018PDU....22..189B} to reproduce the system and simulate the observed images based on parameters fitted by \cite{RXJ_paras}. Then, we use \texttt{photutils} \cite{photutils} for point spread function (PSF) photometry measurement to obtain the observed magnitudes and the corresponding errors. 
Besides, we simulate another four systems, namely HE 0435-1223, SDSS 1206+4332, WFI 2033-4723 and PG 1115+080, by using the public simulation results of $\kappa$, $\gamma$ and $f_{*}$ without lens modeling. 
Here, the $f_\ast$ is the stellar mass fraction, which is relevant parameter for microlensing. 
For these systems, simulated images are no longer generated and the measurement errors are calculated through the signal-to-noise ratio. The corresponding results are shown in the Appendix.

%%%%%%%%%%%%%%%%%%%%%%%%%%%%%%%%%%%%%%%%%%
\subsection{Intrinsic light-curves of quasars}
\label{sec:unlens_lc}

Fluctuations in the intrinsic brightness of quasars are caused by the activity of the accretion disks, which have been found to be well described by Continuous Auto Regressive (CAR) process \cite{2009ApJ...698..895K}. 
% CAR is equivalent to a Gaussian Process where the covariance between two points decreases as a function of their temporal separation,
CAR algorithm allows us to describe quasar light-curves with three free parameters: a characteristic timescale $\tau$ in days, which represents the time required for e-folding reduction in correlation between two points; amplitude of fluctuations $\sigma$ in $\sqrt{\rm mag/day}$; and the mean magnitude of the light-curve in the absence of fluctuation $\bar{M}$. The magnitude of the image at time $t$ can be written as \cite{TDC1}
\begin{equation}
    M(t)=e^{-t / \tau} M(0)+\bar{M}\left(1-e^{-t / \tau}\right)+\sigma \int_{0}^{t} e^{-(t-s) / \tau} d B(s),
	\label{eq:CAR}
\end{equation}
where fluctuations are generated by the integrand. 
$dB(s)$ is a normally distributed value with zero mean and standard deviation $dt$.
%We simulate the intrinsic light-curves of quasars over a decade with $\bar{M}=18,\tau=300,\sigma=0.01$ as shown in Figure~\ref{fig:quasar}. 
%One can see that the typical scatters in magnitude is about $0.3$. 
In the following simulations, we set $\tau=300$ and $\sigma=0.01$, which are the typical values in the CAR model \cite{TDC1}. $\bar{M}$ of each images are set to the observed magnitude from the CASTLES catalogs \footnote{cfa-www.harvard.edu/castles}. 

%%%%%%%%%%%%%%%%%%%%%%%%%%%%%%%%%%%%%%%%%%
%\begin{figure}
%    \includegraphics[width=\columnwidth]{lc_year.pdf}
%    \caption{Example of quasar intrinsic light-curve generated by the Continuous Auto Regressive model.}
%    \label{fig:quasar}
%\end{figure}

%%%%%%%%%%%%%%%%%%%%%%%%%%%%%%%%%%%%%%%%%%
\subsection{Lens modeling of RX J1131}

\begin{figure}
    \centering
    \includegraphics[width=0.6\columnwidth]{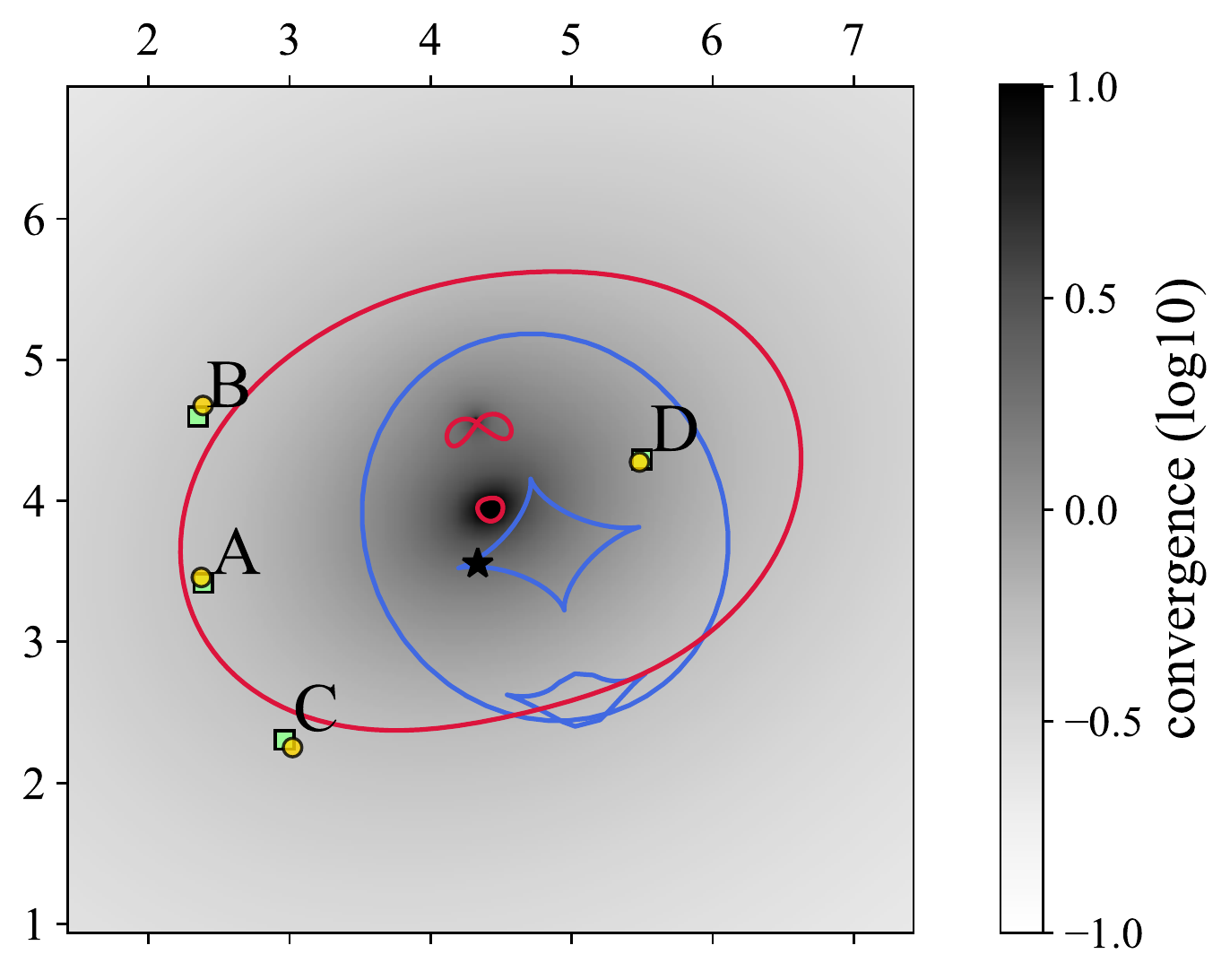}
    \caption{This figure shows the geometry of RX J1131. 
    % The field of view is $6''\times6''$.
    The scale of the view is in the unit of arcsec.
    The green boxes represent the observed positions of the quasar images.
    All other marks in the figure are simulation results.
    The asterisk indicates the position of the quasar, and the yellow circles indicate the simulated positions of the images. 
    The blue and red curves represent the caustics and critical curves respectively.
    The intensity of the gray shadow represents the convergence, and only the intensity between -1 and 1 of logarithmic scale is drawn in the figure.  }
    \label{fig:lensmodel}
\end{figure}

Figure~\ref{fig:lensmodel} shows the geometry of RX J1131, simulated by \texttt{lenstronomy} according to the models and parameters presented in \cite{RXJ_paras}. It is constructed from the power-law mass profiles of the main lens and satellite lens, plus the external shear on the lens plane. The brightness distribution follows the S\'{e}rsic profile.
In the figure, the asterisk indicates the position of the quasar, and the yellow circles indicate the positions of the images. These are from the lens modeling. To demonstrate the accuracy of our simulation, we also plot the observed positions of the images in green boxes. 
The intensity of the gray shadow represents the convergence in logarithmic scale, while the blue and red curves represent the caustics and critical curves, respectively.
One can see that the positions of the images from observation and simulation almost overlap. It means that our simulation can faithfully reproduce the real observation. 

%%%%%%%%%%%%%%%%%%%%%%%%%%%%%%%%%%%%%%%%%%

%%%%%%%%%%%%%%%%%%%%%%%%%%%%%%%%%%%%%%%%%%
\subsection{Microlensing}
\label{sec:microlens}

So far, we have considered a smooth mass distribution for the strong lensing phenomena. Actually, galaxies are constituted by each individual stellar as well as inter-stellar media. 
The mass of lens surface shall be discretized and can be divided into continuous matter and compact matter. 
If the angular Einstein radius of the lens is much smaller than the angular size of the source, and if there is a numerous population, then the surface mass distribution is considered as ``continuous''. For example, the gas and dust particles in the lens galaxies are much smaller than the luminous sources, hence, they are always treated as the continuous component. 
The angular Einstein radius of a star is comparable to the size of quasar, so in this case, stars cannot be regarded as continuous mass but compact mass. However, stars should be regarded as continuous matter when the source is a galaxy \cite{2001stgl.book.....P}.

Microlensing can be thought of strong lensing at a small scale produced by compact matter. 
Because the image separation induced by microlensing is too small to be resolved, one can only hunt for the microlensing effect via the variation in the magnification. 
Here, we denote the total mass surface density as $\kappa$, the continuous mass surface density as $\kappa_{c}$ and the compact mass surface density $\kappa_{s}$. 
Besides, we introduce the stellar fraction $f_{*}$ for lens galaxies, which is the ratio of the stellar mass to the total mass.  
The general lens equation with microlensing reads
\begin{equation}
    \vec{y}=\left(\begin{array}{cc}
    1-\gamma & 0 \\
    0 & 1+\gamma
    \end{array}\right) \vec{x}-\kappa_{c} \vec{x}-\sum_{i=1}^{N_{*}} \frac{m_{i}\left(\vec{x}-\vec{x}_{i}\right)}{\left(\vec{x}-\vec{x}_{i}\right)^{2}}.
    \label{eq:micro}
\end{equation}

\begin{figure}
    \centering
    \includegraphics[width=0.9\columnwidth]{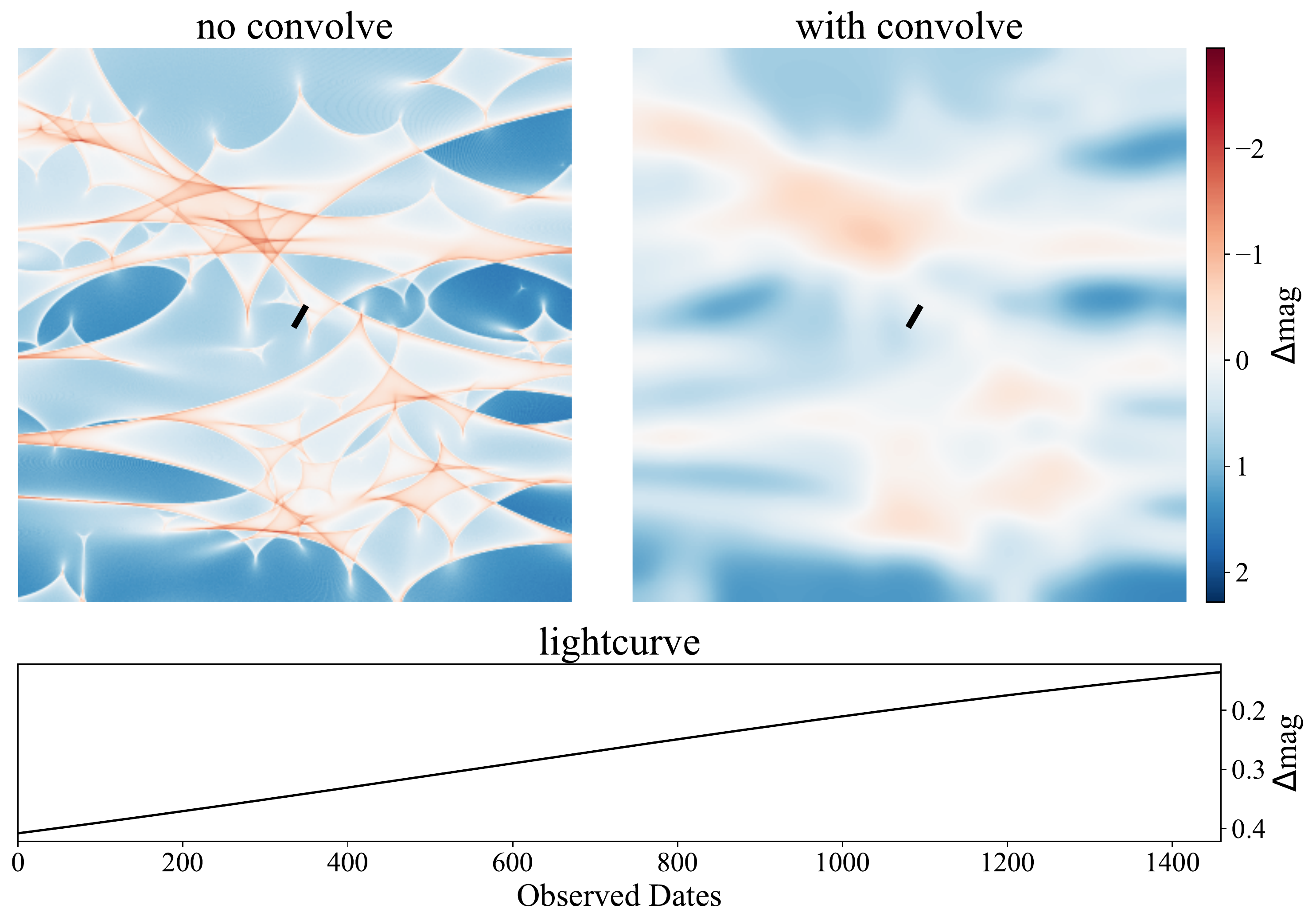}
    \caption{This figure shows the microlensing effects in the images B of RX J1131 over a four-year period, where $\kappa_s=0.194, \kappa_c=0.253, \gamma=0.478$. Here, the magnitude variation is generated by microlensing alone. 
    For RX J1131, $\kappa$ and $\gamma$ are calculated from the lens modeling. $f_{*}$ follows previous measurements which divides $\kappa$ into $\kappa_s$ and $\kappa_c$.
    % For RX J1131, $\kappa$ and $\gamma$ are calculated from the lens modeling. $f_{*}$ follows \cite{Chen2019} which divides $\kappa$ into $\kappa_s$ and $\kappa_c$.
    The left sub-panel is the transverse trajectory (black bar) of the quasar by assuming a point mass source model, labeled as ``no convolve''. The right sub-panel is the more realistic case by considering the finite area effect of the sources, labeled as ``with convolve''. 
    The box size is 4 Einstein radii of a solar mass lens.
    The bottom sub-panel shows the microlensing induced time variation in the case of ``with convolve''.
    }
    \label{fig:microlens}
\end{figure}

In this paper, we use FORTRAN package \texttt{microlens} developed by \cite{1999JCoAM.109..353W} to simulate the magnification maps of microlensing with the method of re-shooting. 
The spatial distribution of the stellar are random. The mass distribution is chosen to follow Salpeter's initial mass distribution function as $d N / d M \propto M^{-2.35}$. The maximum and minimum mass limits are $10 M_{\odot}$ and $0.01 M_{\odot}$, respectively. Given $\kappa_{s}$, $\kappa_{c}$ and shear $\gamma$, we generate magnification maps of images. Figure \ref{fig:microlens} shows the map of image B of RX J1131, in which $\kappa$ and $\gamma$ are calculated from the lens modeling and $f_{*}$ follows \cite{Chen2019} which divides $\kappa$ into $\kappa_s$ and $\kappa_c$.
The sub-panels labeled with ``no convolve'' denote for the case where we treat the quasar as an ideal point source. The ``with convolve'' ones are the more realistic case, where we take the finite quasar source area into consideration. Hence, compared with the left, the caustics in the right sub-plots are more blurred. The convolution is done by smoothing the point source map with a Gaussian radius $R_{\mathrm{src}}=5\times10^{13}$m. 
The black bar in the figure denotes for the transverse trajectory of the quasar in the source plane during the observation period, namely 4 years in this case. We randomly pick up a direction and set the relative velocity between source and lens as $v_{\mathrm{vel}}=500km/s$, which is consistent with the mean velocity calculated by \cite{1131_vel} as $488km/s$.
The box size for these two magnification maps is 4 $\theta_\odot$ (Einstein radii of a solar mass lens). As the redshifts of the lens and source for RX J1131 are 0.295 and 0.657, respectively \cite{2020PyCS}. For RX J1131, $\theta_\odot$ is 2.13 $\mu as$ and $R_{\mathrm{src}}=0.11 \theta_\odot$ and $v_{\mathrm{vel}}=3.44\times10^{-2} \theta_\odot / \mathrm{yr}$.

\begin{figure}
    \centering
    \includegraphics[width=\columnwidth]{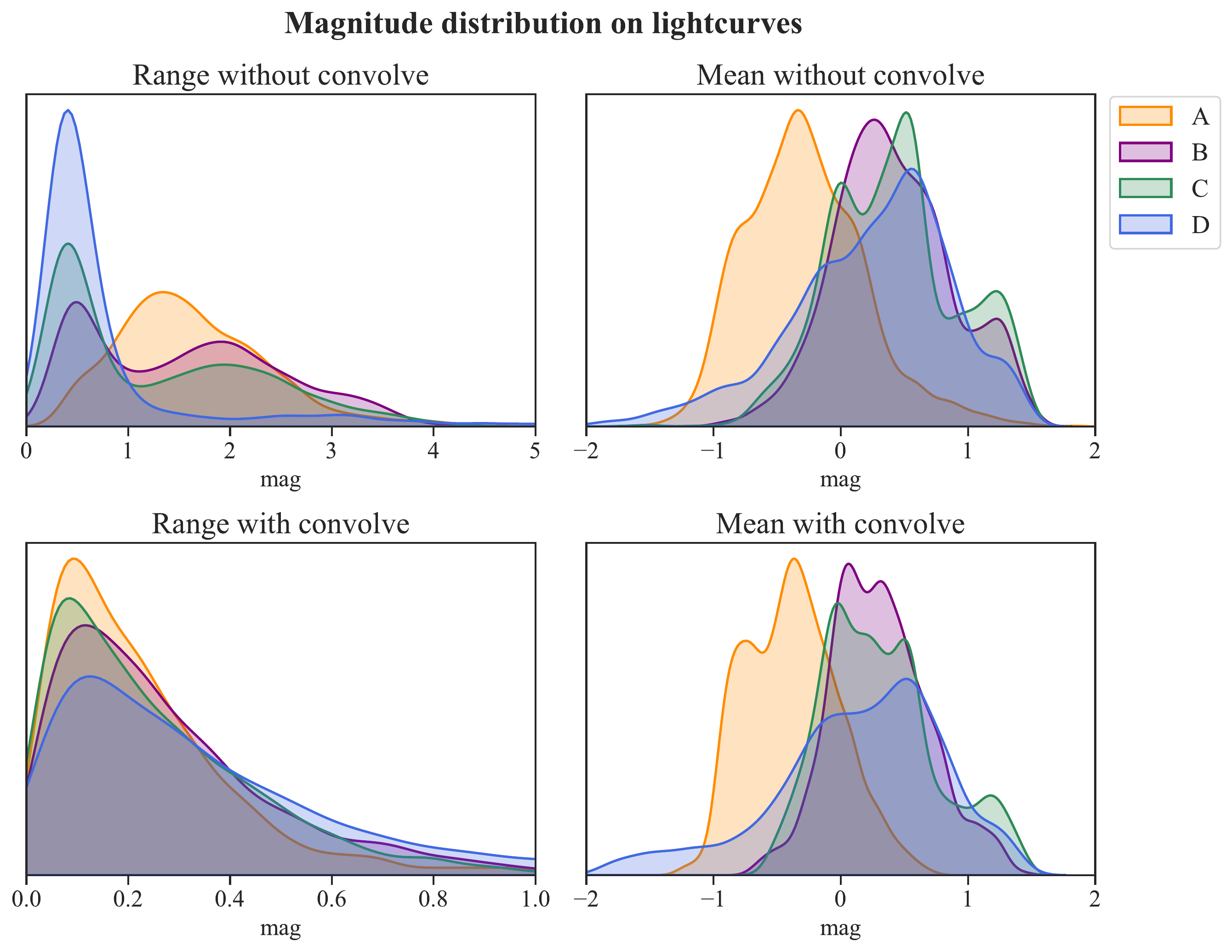}
    \caption{The probability distribution of range and average microlensing effect over the 4-year observation period for the "with convolution" and "without convolution" cases. The statistics are obtained from 5000 randomly distributed transverse trajectories over the 4-year observation period.}
    \label{fig:micro_distribution}
\end{figure}

Figure \ref{fig:micro_distribution} shows the probability distribution of microlensing effect for the "with convolution" and "without convolution" cases. 
We calculate the range and average microlensing effects based on 5000 randomly distributed transverse trajectories over the 4-year observation period under two different cases.
The four images are represented by orange, purple, green and blue, respectively.
It can be seen that the variation range of the microlensing curves decreases significantly after convolution while the average distribution is almost unchanged.  

% Figure \ref{fig:micro-violin} shows the microlensing magnification distributions of the four images of RXJ1131. 
% The histograms add up all the pixels in the zoom-in area of each images. The orange region on the left represents the ``no convolve'' magnification distribution, while the blue region on the right represents the ``with convolve'' magnification distribution.

%%%%%%%%%%%%%%%%%%%%%%%%%%%%%%%%%%%%%%%%%%

%%%%%%%%%%%%%%%%%%%%%%%%%%%%%%%%%%%%%%%%%%

%%%%%%%%%%%%%%%%%%%%%%%%%%%%%%%%%%%%%%%%%%
% \begin{figure}
%     \centering
%     \includegraphics[width=0.9\columnwidth]{micro_violin.pdf}
%     \caption{Magnification distributions from microlensing effects alone on four images of RXJ1131, before (orange on the left) and after (blue on the right) convolution, respectively. All units are magnitudes.}
%     \label{fig:micro-violin}
% \end{figure}

%%%%%%%%%%%%%%%%%%%%%%%%%%%%%%%%%%%%%%%%%%
\subsection{Cadences, campaign lengths and photometric errors}
\label{sec:sampling}

%The location of the observatory determines whether the targets can be observed. The actual situation of the telescope and site, such as bright moonlight and poor weather, should also be considered in the process of observation strategies. 

\begin{figure}
    \centering
    \includegraphics[width=\columnwidth]{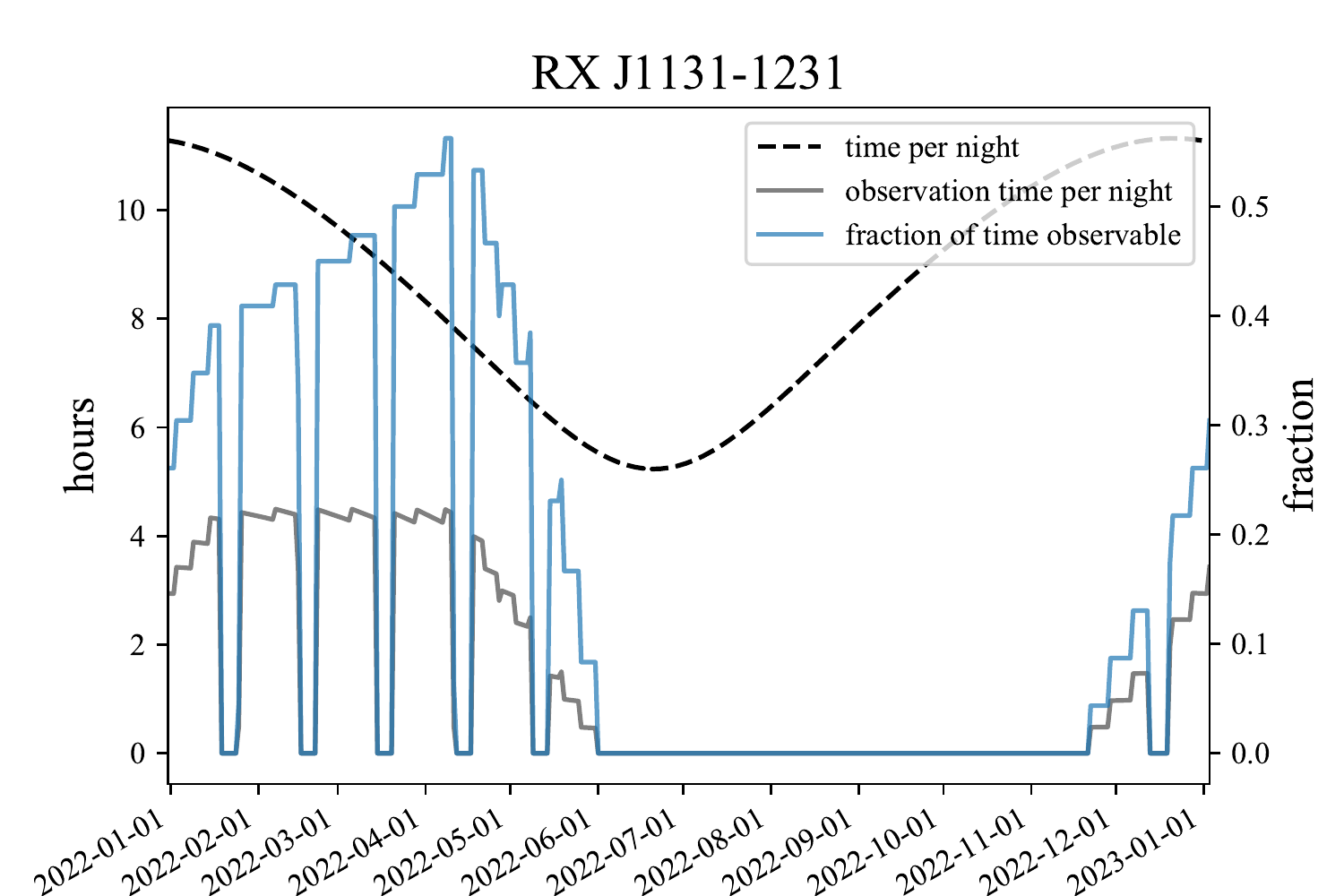}
    \caption{Rough calculation of the observation time of RX J1131 at the Muztagh-Ata site in 2022 under some simple conditions. The observable conditions for the targets are: (1) between 30-80 degrees of altitude (2) at the least 45 degrees away from the moon. The black dashed curve in the figure represents the time between astronomical evening and morning each night (the time when the Sun's altitude is below -18 degrees). The gray curve represents the observable time each night. The units of these two curves are hours. The blue curve represents the proportion of observable time per night to that night. From this calculation, we can conclude that RX J1131 is observable in about 200 days over the year.}
    \label{fig:obs_time}
\end{figure}

RX J1131 is a typical target located within the observable sky patch of Muztagh-Ata 1.93m telescope. We made a rough estimation of its observable time with two conditions: (1) the target shall be in the altitude range 
between 30-80 degrees; (2) the target shall be 45 degrees away from the moon at the least. According to our calculations, the observable time of RX J1131 is about 200 days per year from our site, as shown in Figure \ref{fig:obs_time}.
Our purpose is to discuss the capability of measuring TDSL with Muztagh-Ata 1.93m telescope, rather than to give prediction for a specific target. Based on the field measurements at Muztagh-Ata site \cite{2020RAA....20...86X,2020RAA....20...87X,2020RAA....20...88X}, we believe 200 observable days per season (\textit{ie.}, per year) is reasonable number for our simulations. 
%Although this may not be the actual observable time of the object, for example, WFI2033 cannot be observed here at all. 
In order to demonstrate the robustness of measuring TDSL with 1.93m telescope, here we consider several cadences as well as observation campaign length. In details, we simulate the monitoring with 2 cadences (1 and 3 days) and 4 campaign lengths (2, 3, 4 and 8 seasons).   

%%%%%%%%%%%%%%%%%%%%%%%%%%%%%%%%%%%%%%%%%%

%%%%%%%%%%%%%%%%%%%%%%%%%%%%%%%%%%%%%%%%%%
\begin{figure}
    \centering
    \includegraphics[width=\columnwidth]{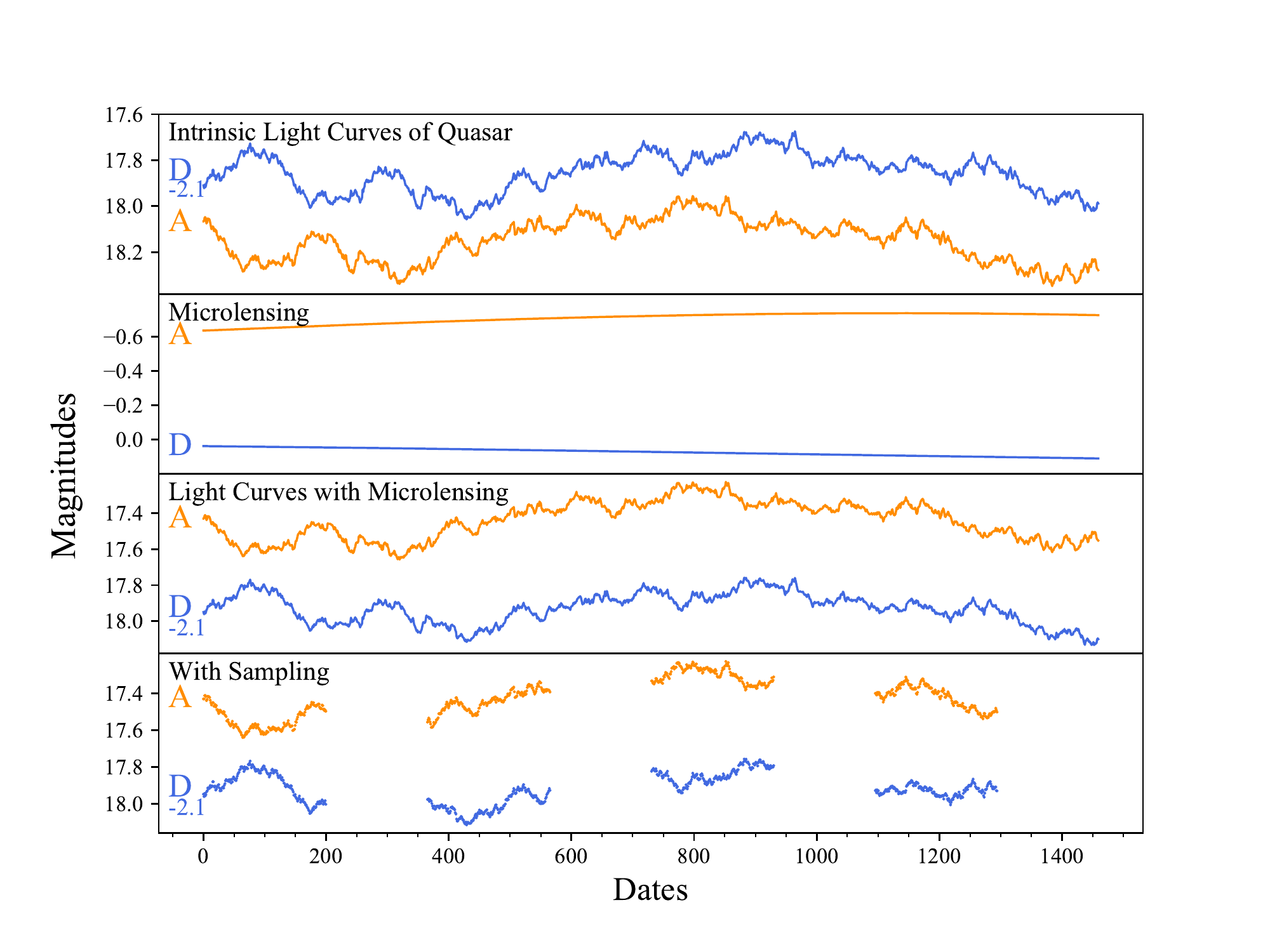}
    \caption{Illustration of the process for generating the brightness of the simulated images.
    Since the time delays between A, B and C are much smaller than that between A and D, in order to make the image clearer, only A and D images are drawn in this figure. Orange denotes for image A and blue for image D. The brightness of image D is reduced by 2.1 magnitudes in the display.
    The panels from top to bottom show: (1) the intrinsic light-curves containing strong lensing contributions (including time delays and brightness variations); (2) the microlensing contributions in magnitudes; (3) quasar light-curves including both the strong lensing and microlensing contributions; (4) the results of down-sampling with a fiducial cadence (1 day) and season length (4 seasons), which are subsequently used for the rest calculation.}
    \label{fig:process}
\end{figure}

Figure \ref{fig:process} illustrates the light-curve data generation process. As an example, we show the 4 season data with 1 day cadence. 
To get a better visualization effect, we only show the light-curves of the images with the maximum delay, namely image A (orange) and D (blue).
To avoid the overlap in the figure, the brightness of image D is subtracted with a number 2.1 in magnitudes.
From the top to bottom, we firstly generate the intrinsic quasar light-curve and plus the time delay and magnification from strong lensing; then add magnification from microlensing on top of the intrinsic light variation\footnote{Here, we only consider the magnification variation due to microlensing, but ignore the time delay induced by microlensing.}; and finally take the cadence effect into account. 
Here we assume the constant sky brightness and seeing 21.35 $\mathrm{mag} \operatorname{arcsec}^{-2}$ and $0.82''$, respectively. And the readout noise is set as $R_{\text{readout}}=5$. 

\begin{figure}
    \includegraphics[width=\columnwidth]{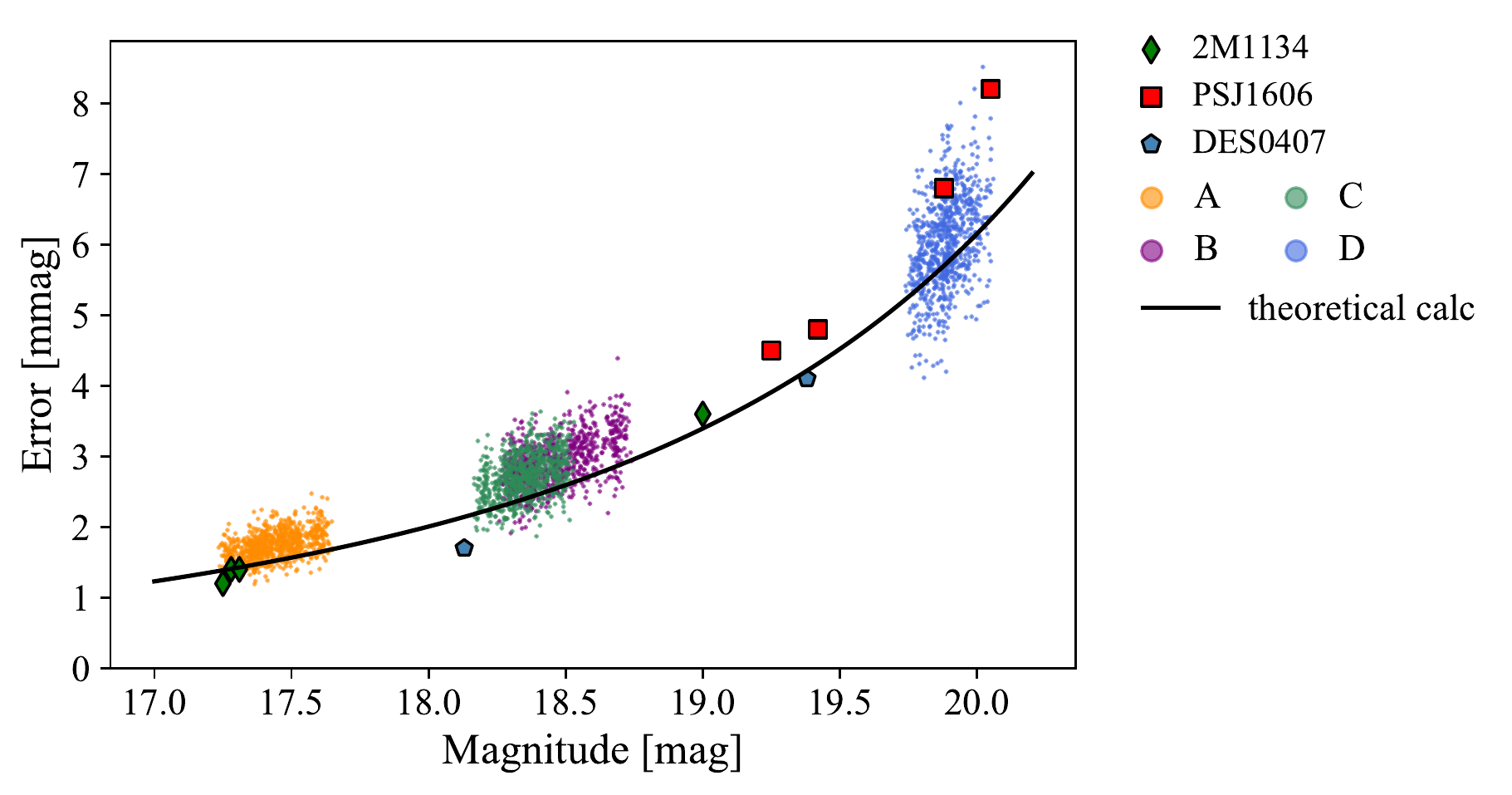}
    \caption{The brightness and photometric errors distribution of the images of RX J1131, 2M 1134, PS J1606 and DES 0407. Orange, purple, green and blue dots represent the four images of RX J1131 from PSF photometry, respectively. They are under the observation strategy of 4-season campaign length and 1-day sampling interval and each dot represents the photometry measurement in one observation night. The black curve is the theoretical calculation with signal-to-noise ratio (SNR) through Eq. (\ref{eq:SNR}) and Eq.(\ref{eq:sigma_SNR}).
    The brightness of 2M 1134, PS J1606, DES 0407 are similar to RX J1131. And the empirical noise $\sigma_{\mathrm{emp}}$ of them are represented in green diamonds, red boxes and blue pentagons in the figure, respectively. }
    \label{fig:mag_err}
\end{figure}

For RX J1131, we generate the brightness data by \texttt{lenstronomy}. The statistical photon errors are estimated via Monte Carlo simulations rather than the analytical formula. In each pixels, the background Gaussian noise per second is generated randomly according to the standard deviation
\begin{equation}
    \sigma_{\mathrm{img}}=\frac{\sqrt{R_{\text{readout}}^{2}+t_{\mathrm{obs}} \cdot n_{\mathrm{sky}}}}{t_{\mathrm{obs}}}\;,
\end{equation}
where $n_{\mathrm{sky}}$ is counts per second per pixel from the sky brightness. The photon fluctuations of the lens images are generated according to the Poisson distribution of the brightness. Each images are obtained by $t_{\mathrm{obs}}=300s$ exposure. We call these maps as ``data maps''. Besides, we generate a ``reference map'' with a much longer exposure time, which is used for subtraction of lens galaxy and host galaxy as their brightness are invariant. 
The brightness used to generate each image is the result of taking into account the light transmission rate of the telescope and the quantum efficiency of the CCD.  
After this, the images are smoothed by the Gaussian PSF with seeing size $0.82''$. 
We get the quasar images by subtracting the ``data maps'' from the ``reference map'' and then calculate the relative brightness via PSF photometry by \texttt{photutils}. 
At each epoch, 50 realisations of the data maps are generated, and the PSF photometric errors on the lensed images are calculated according to Eq.(\ref{eq:error}), where $m_i$ is the brightness of image relative to the reference map and $\bar{m}$ is the mean value of $m_i$
\begin{equation}
    \sigma^{2}=\frac{1}{N-1} \sum_{i=1}^{N}\left(m_{i}-\overline{m}\right)^{2}.
    \label{eq:error}
\end{equation}
Figure \ref{fig:mag_err} shows the brightness and photometric errors distribution of the images of RX J1131 in this work and several previous observations in TDCOSMO project \cite{TDCOSMO2}.
The orange, purple, green and blue dots represent the results of 4 images of RX J1131 from PSF photometry, respectively. They are under the observation strategy of 4-season campaign length and 1-day sampling interval. Each dot represents the photometry measurement in one observation night. The black curve is the theoretical calculation with signal-to-noise ratio (SNR) through Eq. (\ref{eq:SNR}) and Eq.(\ref{eq:sigma_SNR}).
One can see that the error from theoretical calculation can be taken as a face value. 
2M 1134-2103, PS J1606-2333, DES 0407-5006 are three systems with similar brightness to RX J1131. 
The green diamonds, red boxes and blue pentagons in the figure represent the empirical noise $\sigma_{\mathrm{emp}}$, corresponding to the standard deviation of the measured image flux for these three systems, respectively.
It can be seen that our photometric errors are roughly at the same level as TDCOSMO project.
% One can see that the photometry errors are a few milli-magnitude for the source with magnitude of 18. This is roughly at the same level as TDCOSMO project \cite{TDCOSMO2}. 

%%%%%%%%%%%%%%%%%%%%%%%%%%%%%%%%%%%%%%%%%%

%%%%%%%%%%%%%%%%%%%%%%%%%%%%%%%%%%
\section{Time delay measurement}
\label{sec:measure}

In this section, we will present the time delay measurements of RX J1131 based the aforementioned simulated 4 seasons light-curve data. The brightness and corresponding errors are measured as illustrated in the Section \ref{sec:sampling}. 
%\shu{It is unclear which light-curves are used, the ones directly from the simulation or light-curves measured from single-epoch photometry? }.
We use \texttt{PyCS}, a publicly available python toolbox developed by the COSMOGRAIL collaboration~\cite{PyCS3}. It bases on the iterative nonlinear optimization algorithms and is fully data-driven. It can simultaneously estimate both the intrinsic time delay as well as those induced by microlensing. A free-knot spline estimator and a regression-difference estimator are provided. Both methods perform well in terms of precision and accuracy in the Time-Delay Challenge \cite{TDC2}. Nowadays, time delay cosmography studies are mostly based on these two methods, such as H0LiCOW and COSMOGRAIL collaborations.
In this work, we use the free-knot spline estimator to measure the time delay between images and evaluate the uncertainties.

The free-knot spline estimator models light-curves with analytical
spline functions. 
% The position of the knots are free parameters in the fitting procedure. 
The details of the algorithm are as what follows. We go first for a rough estimation of the time-delay and then refine it for adjusting the local features. The position of the knots are fixed in the rough estimation. This step is used to search the global solution, which is less sensitive to the fine structures. After that, we freely vary the knots within the range of 10 days for finding the local features.
A single common spline fits simultaneously for the intrinsic variations of all images. 
And independent splines fit individually on each light-curves for microlensing. The parameters of the estimator are: the initial spacing between the knots of the intrinsic spline ($\eta$) and the initial spacing between the knots of the extrinsic splines ($\eta_{ml}$). They represent the mean spacing between knots before starting the optimisation.

\begin{figure*}
    \centering
    \subfigure[Spline fit of data and errors generated in Section~\ref{sec:sampling}. The black curve represents the intrinsic quasar light variation shared by all images ($\eta=35$ days). Orange, purple, green and blue curves fit the extrinsic variations of the four images, respectively ($\eta_{ml}=150$ days). The horizontal dashed lines are the median of the extrinsic variations of four images. The data points are shifted in time by the best time delay measurements, and corrected from their modelled extrinsic variations.
    Note that vertical ticks do not represent error bars, but are used to visually indicate the positions of knots. ]{
        \includegraphics[width=\columnwidth]{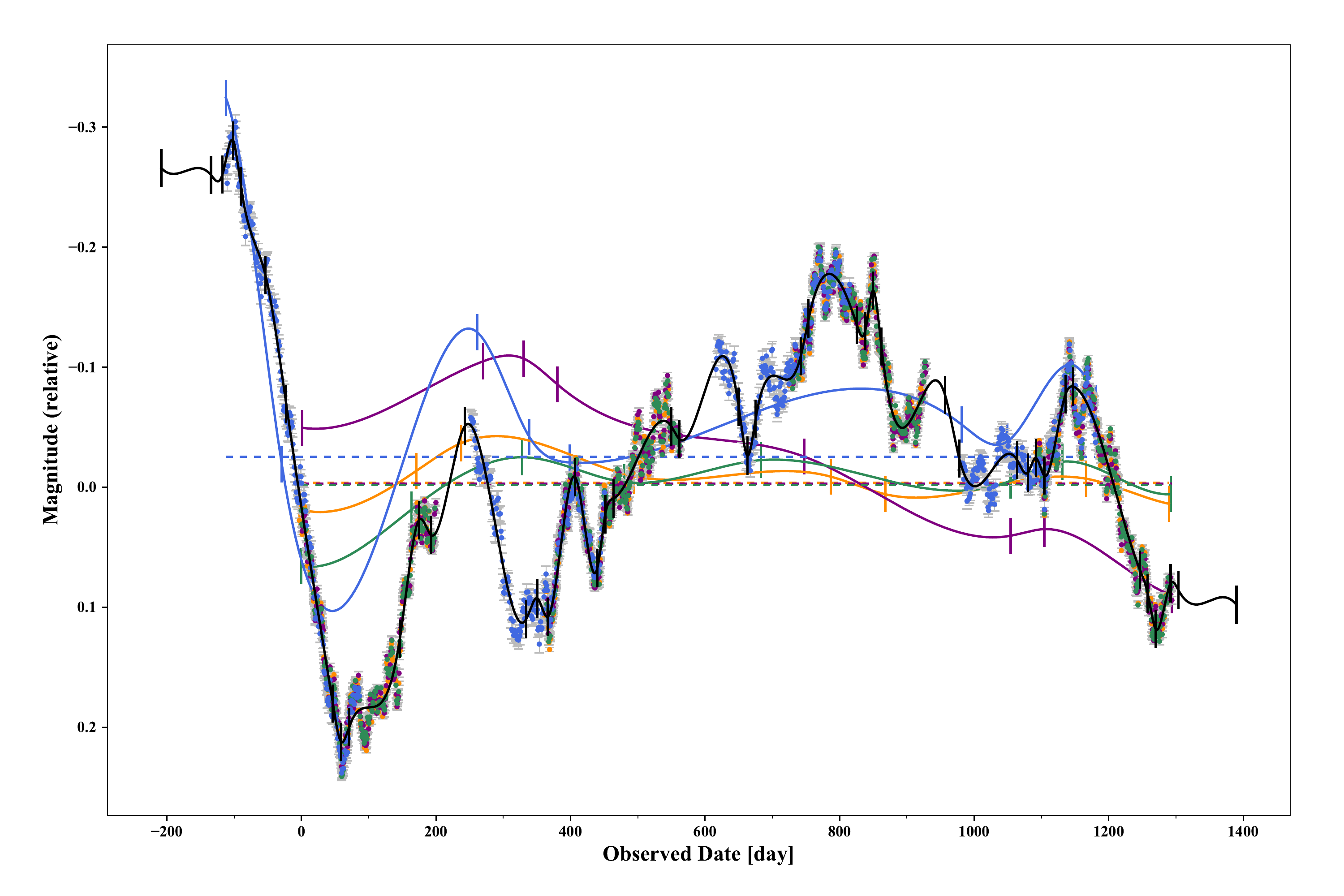}
        \label{subfig:spline}
        }
    \quad
    \subfigure[Residuals of the spline fit. The horizontal solid curves represent for the median absolute deviation.]{
        \includegraphics[width=0.9\columnwidth]{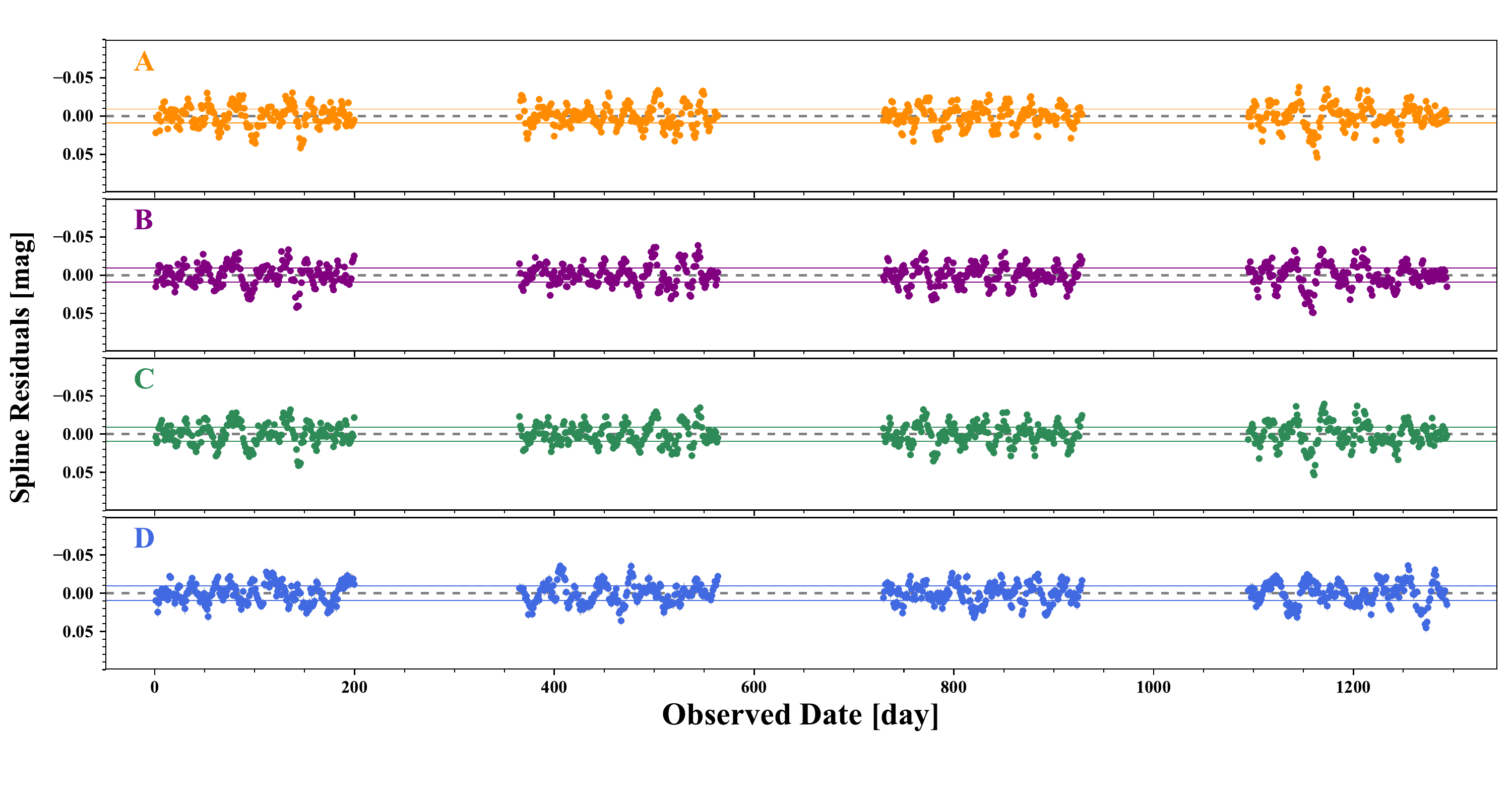}
        \label{subfig:res}
        }
    \caption{Fitting result of the light-curves by using the free-knot spline estimator. Figure \ref{subfig:spline} shows the fitting splines and curves while figure \ref{subfig:res} shows the residuals of fitting. The start time of observation starts from day 0.}
    \label{fig:combine}
\end{figure*}

% For each of the parameter pairs ($\eta,\eta_{ml}$), the fitting procedures are divided into rough and fine stages. The rough fitting procedure is used to search the global solution which is less sensitive to the initial guess. After that, the fine fitting procedure is used to search the local features within 10 days. Each iteration step is composed by one rough fitting and one fine fitting. For finding the accurate time delay, we iterate this procedures a few hundred times. We will describe the details in the next paragraph. 
In Figure.~\ref{fig:combine}, we show an intermediate spline fitting result with ($\eta=35~{\rm days},\eta_{ml}=150~{\rm days}$). 
The black curve is the fitting result for the intrinsic light-curve. Orange, purple, green and blue curves are the fitting results for the extrinsic  microlensing effect in image A, B, C and D, respectively\footnote{Due to the code convention, here we show the opposite brightness variation induced by microlensing}. 
In the upper panel, we plot the relative magnitude variation with respect to the reference magnitude. One can see that the typical variation is about $0.1$ mag. The bottom panels are the residuals between the input data and fitting results. The horizontal solid curves represent for the median absolute deviation, which is about $0.01$ mag. Hence, we can conclude that this method gives relatively faithful reconstruction of the original signal.

%%%%%%%%%%%%%%%%%%%%%%%%%
% \begin{figure}
%     \centering
%     \includegraphics[width=\columnwidth]{fall.pdf}
%     \caption{Fitting result of the light-curves by using the free-knot spline estimator. \textit{Top panel:} Spline fit of data and errors generated in Section~\ref{sec:sampling}. The black curve represents the intrinsic quasar light variation shared by all images ($\eta=35$ days). Orange, blue, green and purple curves fit the extrinsic variations of the four images, respectively ($\eta_{ml}=150$ days). The data points are shifted in time by the best time delay measurements.
%     Note that vertical ticks do not represent error bars, but are used to visually indicate the positions of knots.  
%     \textit{Bottom Panel:} Residuals of the spline fit.}
%     \label{fig:spline}
% \end{figure}

The time delay and its errors are estimated as what follows \cite{2020PyCS}. 
(1) We use the first guess delay as the starting point to get the preliminary measurement of time delays, splines and residuals of the fit. The first guess can be the result from the algorithm mentioned above, or simply a visual estimate. (2) In order to reduce the starting point dependence, the input light-curve is measured 500 times with starting points randomly selected around the first guess. The final time delay result is the median value of these measurements. (3) To estimate the uncertainties, we apply the estimator on mock light-curves with basically the same quality as the ``real'' data but different true delays. The signals in the mock light-curves are constructed with the splines instead of the data generation process presented in Section \ref{sec:simulation}. The noise is generated according to the noise power spectrum obtained from the fitting residuals. The splines and residuals used here are from the estimation in step 1.
By shifting these curves in time, 800 sets of curves with known time delays are obtained. By comparing the measured value with the true delay, the uncertainty can be calculated as an orthogonal combination of the worst random error and the worst systematic error.

However, the selection of parameters ($\eta,\eta_{ml}$) will affect the time delay measurements to some extent. If the initial spacing was too large, some fast variations will be missed. A too small initial spacing between knots leads to an over-fitting of the data and also affect the results. Therefore, the choice of $\eta$ and $\eta_{ml}$ must be adapted to the data quality, which mainly depends on the cadence, photometric noise, time scale of data variation, {\it etc}. As stated previously, this method is fully driven by the observed data. It is almost impossible to determine which set of ($\eta, \eta_{ml}$) are the most appropriate one.  

\begin{figure*}
    \centering
    \includegraphics[width=\textwidth]{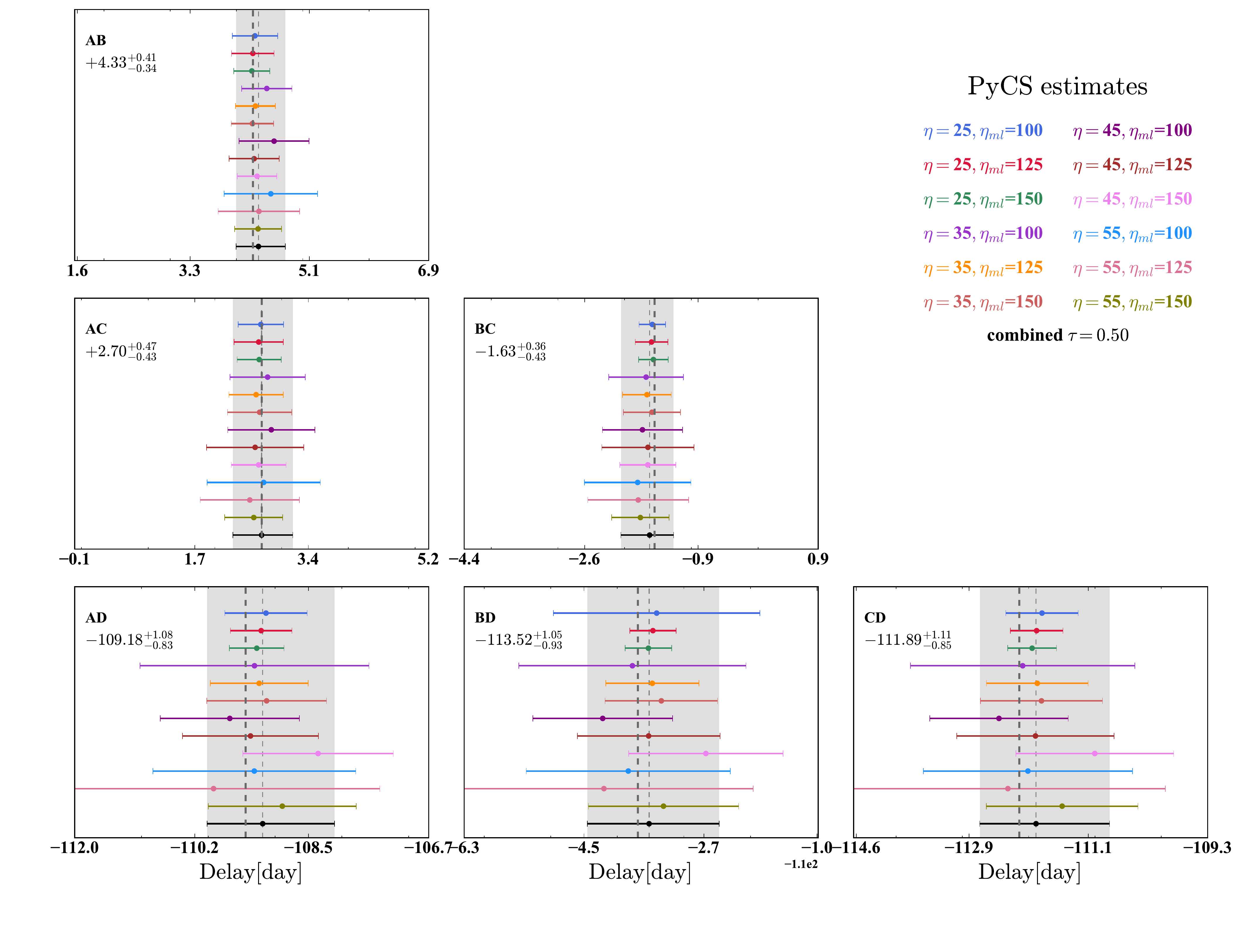}
    \caption{Series of time-delay estimation with different parameters and their combination. 
    Each time-delay estimation shown in the figure correspond to a particular choice of parameters, namely the mean spacing between the knots of the intrinsic spline $\eta$, and of the extrinsic splines $\eta_{ml}$. The combined estimation which corresponds to a threshold of $\tau_{\rm thresh}=0.5$ is shown in black at the bottom, and its uncertainty is indicated as gray shaded band. Dark gray vertical dashed line represents the true time delay. }
    \label{fig:margin}
\end{figure*}

\begin{figure*}
    \centering
    \includegraphics[width=\textwidth]{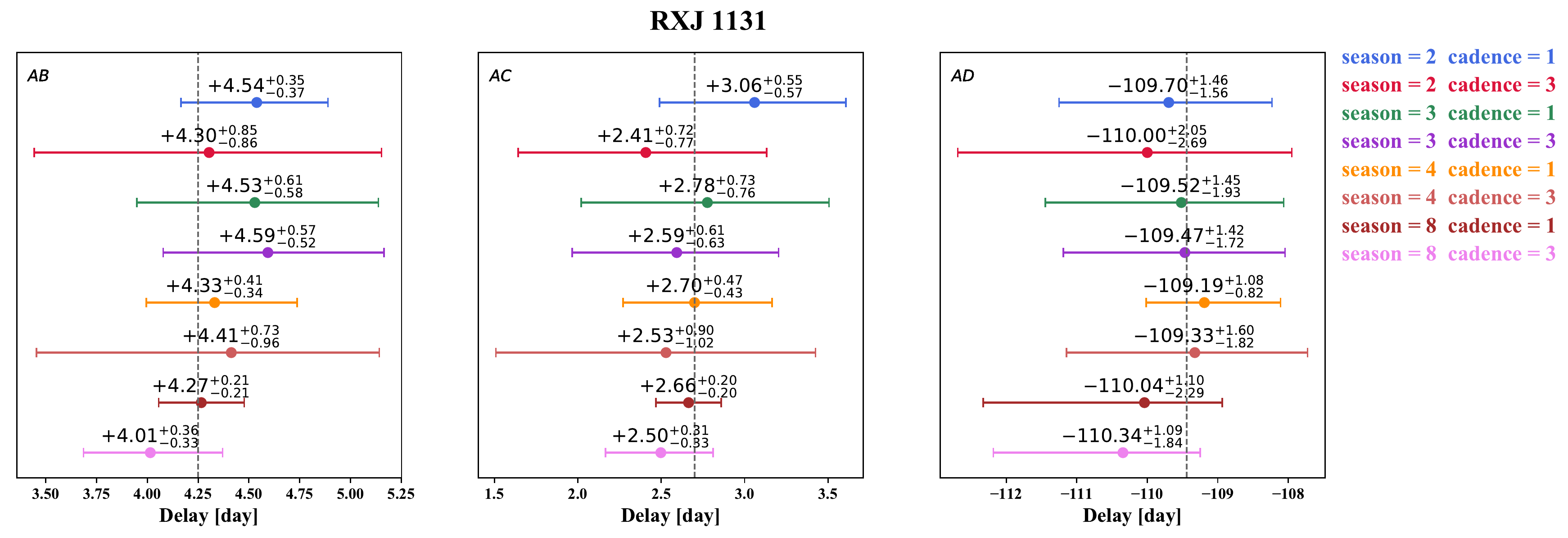}
    \caption{Series of time-delay measurements relative to image A with different observation strategies. Each time-delay estimation shown in the figure correspond to a particular choice of observation strategies, namely the campaign length (season) and sampling interval (cadence). 
    The results are given from the combined time delay measurement, with the algorithm described in the text. The combination threshold parameter is chosen as $\tau_{\rm thresh}=0.5$. The true time delays are shown with gray vertical dashed curves in the figure. }
    \label{fig:RXJ1131}
\end{figure*}

To mitigate this issue, multiple sets of ($\eta, \eta_{ml}$) parameters are measured in the optimal or marginalized sense.
The first version of \texttt{PyCS} was implemented without functioning the multi-parameter combination \cite{2013PyCS}, but was later refined by \cite{2020PyCS}. The improved version adopted an hybrid approach between optimisation and marginalisation. This algorithm marginalises only the sets that do not have significant deviations in the measurements. This deviation, defined by the parameter $\tau$, describes the tension between the set to be marginalized with the reference set \cite{tau_thresh}. If the tension exceeds a certain threshold $\tau_{\mathrm{thresh}}$, we combine the most discrepant estimation with the reference. This combined estimation becomes the new reference and we repeated this process until no further tension exceeds $\tau_{\mathrm{thresh}}$.
Figure.~\ref{fig:margin} illustrates each of ($\eta, \eta_{ml}$) set result, and compare them with the combined estimation. The combined time delay measurement (gray shaded region) is shown in the upper left of each panel, which is consistent with the true delay. 
For a pair of images A and B, a negative value of $\Delta t_{AB}$ means that image A varies first, and the vice versa. 
Results of observations with different cadence and campaign lengths are shown in Figure.~\ref{fig:RXJ1131}. 
One can see that the measurement precision with high cadence are better than those with low cadence. In the former cases, we can achieve the time delay measurement error at the 0.5 day level with 4 seasons campaign length. This is our major result of this paper. 

%%%%%%%%%%%%%%%%%%%%%%%%%

%%%%%%%%%%%%%%%%%%%%%%%%%

%%%%%%%%%%%%%%%%%%%%%%%%%
\section{Conclusions}
\label{sec:conclusion}

Muztagh-Ata site is one of the best astronomical sites all over the world. 
The seeing median value is 0.82 arcsec.  
The median value of the sky brightness is 21.35 $\mathrm{mag} \operatorname{arcsec}^{-2}$ in V-band during the nighttime. For the case without moon, this number can be upgraded into 21.74 $\mathrm{mag}\operatorname{arcsec}^{-2}$ (V-band).  
An effective aperture 1.93m reflector telescope currently is under the construction phase leaded by Beijing Normal University in China. This telescope is equipped with both a three-channel imager/photometer (wavelength covers $3500-11000$ Angstrom) and a low-medium resolution ($\delta\lambda/\lambda=500/2000/7500$) spectrograph. The field of view is $20$ arcmin with the help of the correction mirror. The 300s exposure $10\sigma$ limiting magnitude in V-band is $23.79$. All these numbers indicate the 1.93m telescope is an ideal telescope for monitoring the light variation of the lensed quasar system.  

Based on the observation conditions of the Muztagh-Ata site and the instrument parameters of the 1.93m telescope, we simulate the lensed quasar observations with different cadences and campaign lengths, and forecast the precision of the measured time delay. 
We model quasar intrinsic light-curves, microlensing effect as well as the PSF photometric errors. 
We simulate RX J1131 with lens modeling in the main body section and other four systems without lens modeling in the Appendix. 
According to simulations, for RX J1131-like systems (wide variation in time delay between images) the time-delay observations of strongly lensed quasars can be achieved with the typical precision about $\Delta t=0.5$ day with 4 seasons observation and 1 day cadence. This precision is comparable to the up-coming TDCOSMO project~\cite{Millon:2019slk}. 

This paper presents a preliminary study of the time delay for strongly lensed quasars with Muztagh-Ata 1.93m telescope. The sky brightness and seeing are considered as constant, a bit ideally. Some of the sub-leading order systematics, such as the position of planets and moon, air humidity and weather has not yet been taken into account. We studied the time delay measurement precision with different strategies, which are characterized by the campaign lengths and sampling intervals. When increasing the cadence to one day, we are able to reach a very precise measurement of the time delay in a short campaign rather than decades of observations. As a result, the capability of 1.93m telescope allows it to join the network of TDSL observatories. It will enrich the database for strongly lensed quasar observations and make more precise measurements of time delays. We believe it will help resolve the Hubble tension.

%%%%%%%%%%%%%%%%%%%%%%%%%%%%%%%%%%%%%%%%%%%%%%%%%%
\section*{Data Availability}

The inclusion of a Data Availability Statement is a requirement for articles published in MNRAS. Data Availability Statements provide a standardised format for readers to understand the availability of data underlying the research results described in the article. The statement may refer to original data generated in the course of the study or to third-party data analysed in the article. The statement should describe and provide means of access, where possible, by linking to the data or providing the required accession numbers for the relevant databases or DOIs.

%%%%%%%%%%%%%%%%%%%%%%%%%%%%%%%%%%%%%%%%%%%%%%%%%%
\section*{Acknowledgements}
We thank Sherry Suyu and Frederic Courbin for discussion and comments. 
This work is supported by the China Manned Space Project with No.CMS-CSST-2021-A12, the National Natural Science Foundation of China Grants No. 11973016, No. U2031209, No. 11873006, No. U1931210. YS acknowledges support from the Max Planck Society and the Alexander von Humboldt Foundation in the framework of the Max Planck-Humboldt Research Award endowed by the Federal Ministry of Education and Research.



\bibliographystyle{mnras}

\bibliography{1m9} % if your bibtex file is called 1m9.bib

%%%%%%%%%%%%%%%%%%%%%%%%%%%%%%%%%%%%%%%%%%%%%%%%%%

%%%%%%%%%%%%%%%%% APPENDICES %%%%%%%%%%%%%%%%%%%%%

\appendix
\label{sec:app}

\section{Another four systems}

\begin{figure*}
    \centering
    \includegraphics[width=0.8\textwidth]{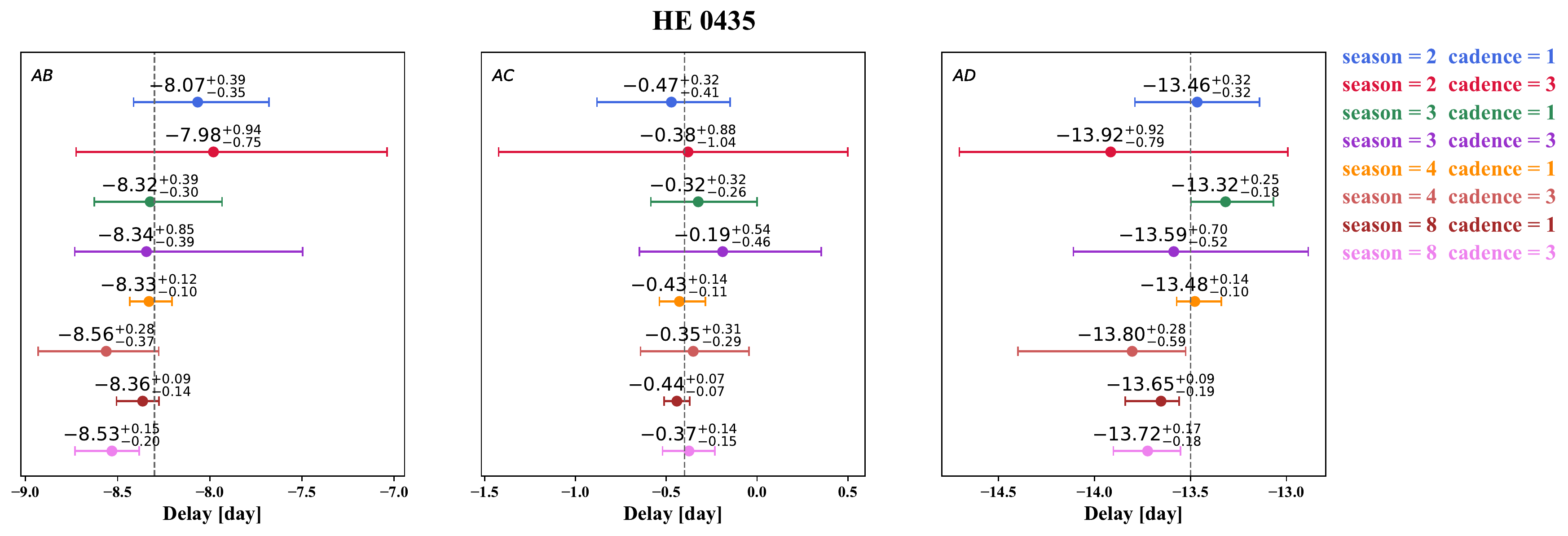}
    \caption{Series of time-delay measurements of HE 0435 relative to image A with different observation strategies. The notations are the same as Figure \ref{fig:RXJ1131}. }
    \label{fig:HE0435}
\end{figure*}

\begin{figure*}
    \centering
    \includegraphics[width=0.8\textwidth]{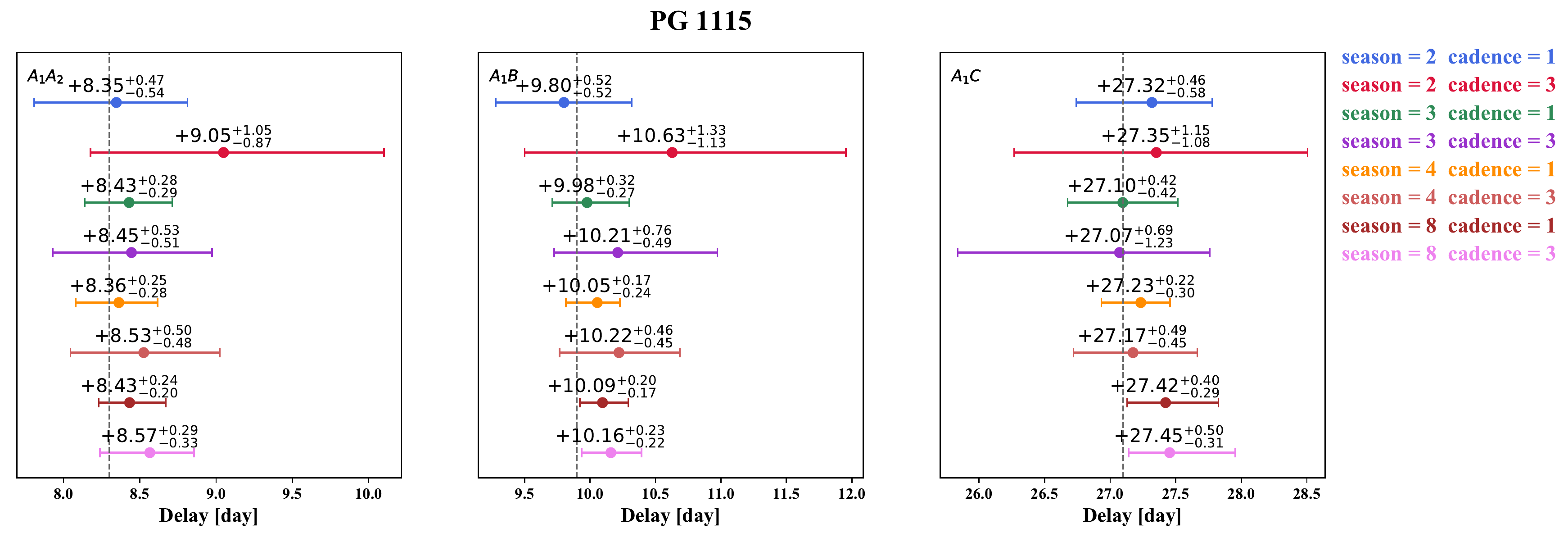}
    \caption{Series of time-delay measurements of PG 1115 relative to image $\mathrm{A_1}$ with different observation strategies. The notations are the same as Figure \ref{fig:RXJ1131}.}
    \label{fig:PG1115}
\end{figure*}

\begin{figure*}
    \centering
    \includegraphics[width=0.8\textwidth]{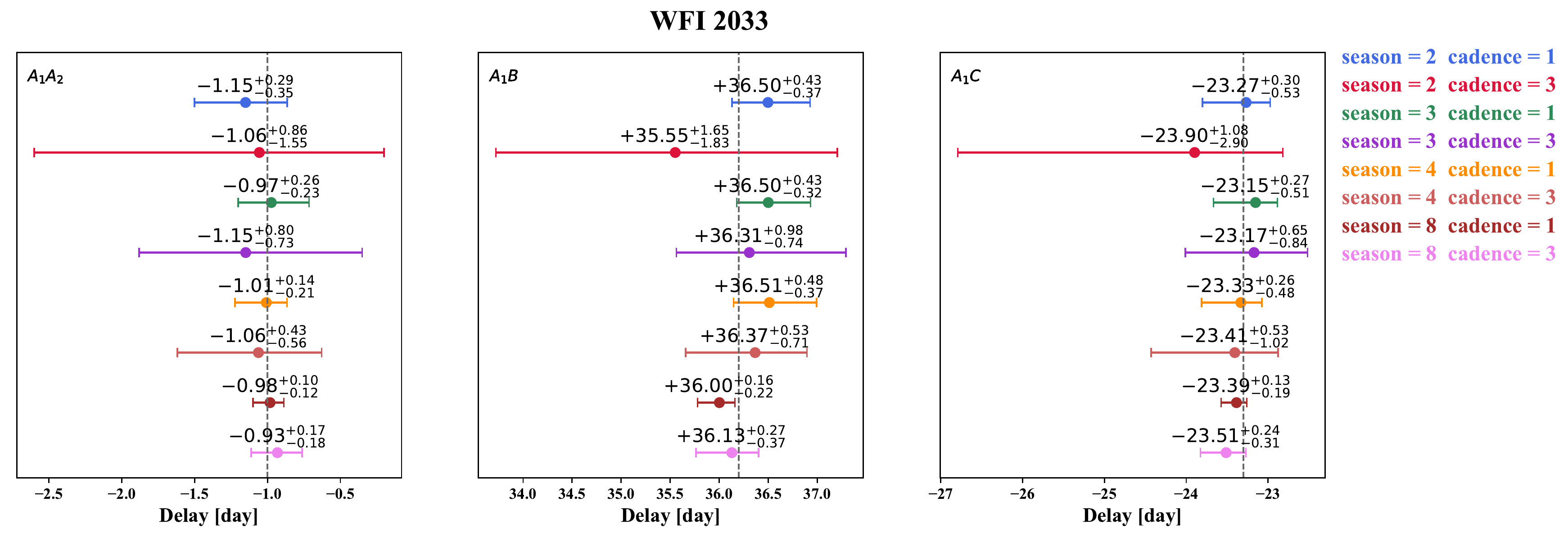}
    \caption{Series of time-delay measurements of WFI 2033 relative to image $\mathrm{A_1}$ with different observation strategies. The notations are the same as Figure \ref{fig:RXJ1131}.}
    \label{fig:WFI2033}
\end{figure*}

\begin{figure*}
    \centering
    \includegraphics[width=0.4\textwidth]{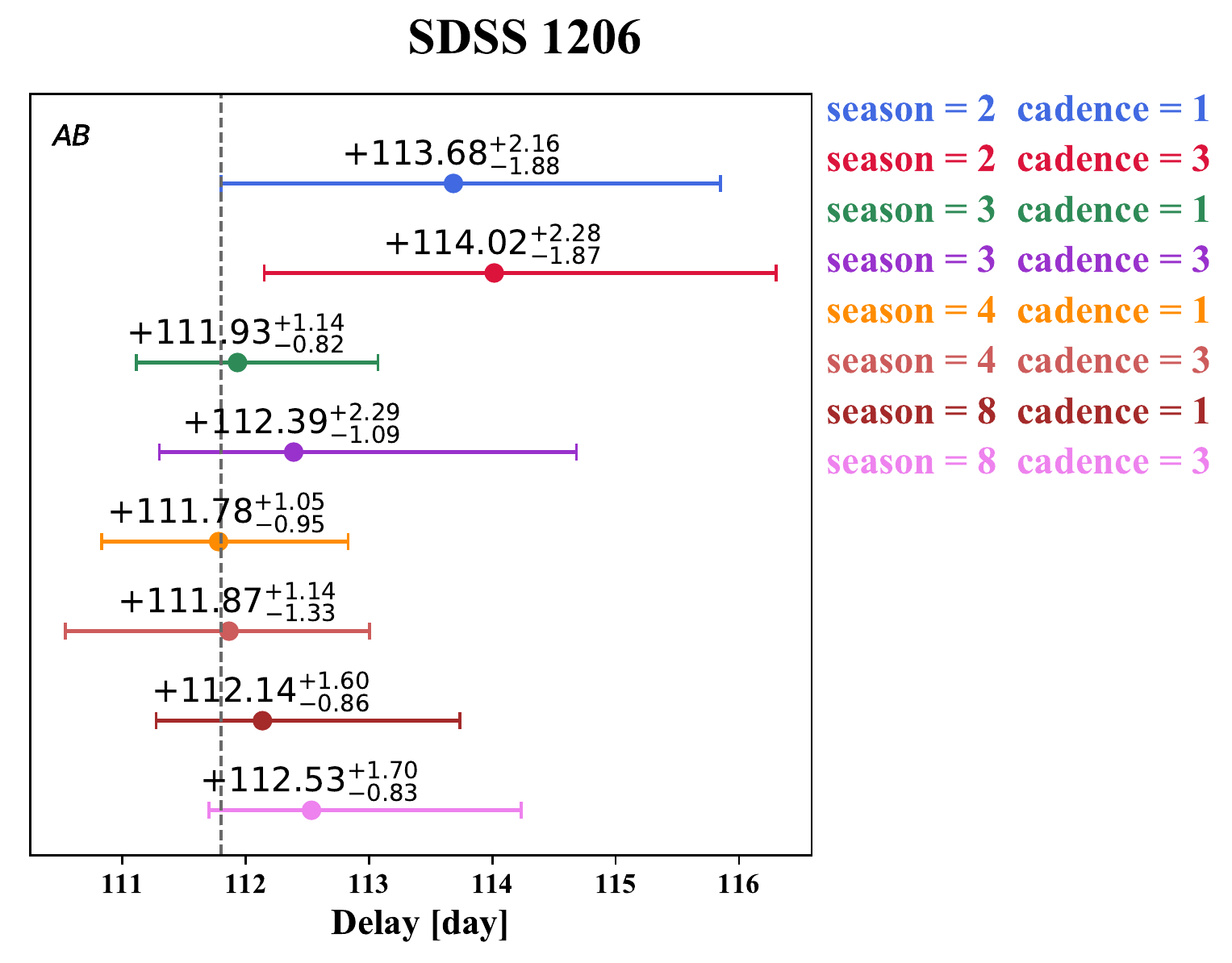}
    \caption{Series of time-delay measurements of SDSS 1206 relative to image A with different observation strategies. The notations are the same as Figure \ref{fig:RXJ1131}.}
    \label{fig:SDSS1206}
\end{figure*}

In the appendix, we show the results of the another four strongly lensed quasar systems, namely HE 0435-1223, SDSS 1206+4332, WFI 2033-4723 and PG 1115+080. 
The time delays of these systems do not differ much from each other.
Since we are lack of the sufficient information to model these 4 systems, we directly use the publicly available simulation results of $\kappa$, $\gamma$ and $f_{*}$ of them. In this case, we do not need to generate the lens images and can directly simulate the light-curves. Instead of using the PSF photometry, here we use the analytical method to calculate the photometric magnitudes and errors. 
The generation of intrinsic light-curves follows Section \ref{sec:unlens_lc} with CAR process. $\tau$ and $\sigma$ are set to 300 and 0.01, respectively. $\bar{M}$ of each image corresponds to observed magnitudes from CASTLES and Gaia\footnote{The brightness of SDSS 1206 is missing in CASTLES, so we use that of Gaia.}.
We adopt previous measurements \cite{2020PyCS,tau_thresh,COSMOXVIII,H0Li_9} as true delays of these light-curves. 
Several important parameters ($\kappa, \gamma, f_{*}$) for microlensing magnification maps are listed in Table~\ref{tab:4sys_paras} and are used to generate microlensing effect as \ref{sec:microlens}.
The photometric error is calculated through the signal-to-noise ratio (SNR), which is defined as
\begin{equation}
    \begin{aligned}
    {\rm SNR} &=\frac{N_{\text {star}}}{\sqrt{N_{\text {star}}+n_{\text {pix}}\left(N_{\text {sky}}+R_{\text {readout}}^{2}\right)}} \\
    &=\frac{{n_{\text {star}} t_{\text {obs}}}}{\sqrt{{n_{\text {star} } t_{\text {obs}}+n_{\text {pix}}\left(n_{\text {sky}} t_{\text {obs}}+R_{\text {readout}}^{2}\right)}}}
    \end{aligned}.
    \label{eq:SNR}
\end{equation}
Then, we convert it into the photometric error in magnitudes \cite{SNR_eq}
\begin{equation}
    \sigma=1.0857/{\rm SNR},
    \label{eq:sigma_SNR}
\end{equation}
where $n_{\text{pix}}$ is the number of pixels covered by the image, mainly determined by seeing. The value of 1.0857 is the correction term between an error in flux (electrons) and that same error in magnitudes. We normally distribute random values with zero mean and standard deviation from Eq. (\ref{eq:sigma_SNR}) onto the brightness images. 
Frankly speaking, the error given in Eq. (\ref{eq:sigma_SNR}) is a bit optimistic and one can take this number as a face value. As a forecast paper, this number sets the upper limit of the photometry measurement of the Muztagh-Ata 1.93m telescope. 
Results of different observation strategies are shown in Figure \ref{fig:HE0435}, \ref{fig:PG1115}, \ref{fig:WFI2033} and \ref{fig:SDSS1206}. Table \ref{tab:results} shows the time delay measurements under the observation strategy of 3-season campaign length. The time delay measurements for these four systems can achieve good accuracy under this observation strategy.

%%%%%%%%%%%%%%%%%%%%%%%%%%%%%%%%%%%%%%%%%%%%%%%%%%
\begin{table*}
    \centering
    \caption{Lensing parameters for creating the microlensing magnification maps.}
    \begin{tabular}{cccccc}
        \hline 
        \text { Name } ($z_l, z_s$) & \text{img} & $\kappa$ & $\gamma$ & $f_{*}$ & \text{Reference} \\
        \hline 
        \text { HE 0435-1223 } & \text{A} & 0.473 & 0.358 & 0.347 & \cite{Chen2019} \\
        (0.454, 1.693)& \text{B} & 0.630 & 0.540 & 0.361 \\
        & \text{C} & 0.494 & 0.327 & 0.334 \\
        & \text{D} & 0.686 & 0.575 & 0.380 \\
        \text { PG 1115+080 } & $\mathrm{A_1}$ & 0.424 & 0.491 & 0.259 & \cite{Chen2019} \\
        (0.311, 1.722)& $\mathrm{A_2}$ & 0.451 & 0.626 & 0.263 \\
        & \text{B} & 0.502 & 0.811 & 0.331 \\
        & \text{C} & 0.356 & 0.315 & 0.203 \\
        \text { WFI 2033-4723 } & $\mathrm{A_1}$ & 0.350 & 0.340 & 0.612 & \cite{COSMOXVIII} \\
        (0.658, 1.662)& $\mathrm{A_2}$ & 0.462 & 0.424 & 0.690 \\
        & \text{B} & 0.281 & 0.309 & 0.519 \\
        & \text{C} & 0.567 & 0.547 & 0.698 \\
        \text{SDSS 1206+4332} & \text{A} & 0.650 & 0.660 & 0.146 & \cite{H0liCOWIV} \\
        (0.748, 1.789)& \text{B} & 0.430 & 0.350 & 0.047 \\
        \hline
    \end{tabular}
    \label{tab:4sys_paras}
\end{table*}

%%%%%%%%%%%%%%%%%%%%%%%%%%%%%%%%%%%%%%%%%%%%%%%%%%
\begin{table*}
    \renewcommand\arraystretch{1.5}
    \centering
    \caption{Time-delay measurements relative to the first image. Under the observation strategy of 3-season campaign length and 1-day or 3-day sampling interval.}
    \begin{tabular}{lll}
        \hline 
        \text   & season = 3 cadence = 1 & season = 3 cadence = 3 \\
        \hline 
        \text { HE 0435-1223 } & \text{AB} = $-8.32^{+0.39}_{-0.30}$ & \text{AB} = $-8.34^{+0.85}_{-0.39}$ \\
        &  \text{AC} = $-0.32^{+0.32}_{-0.26}$ & \text{AC} = $-0.19^{+0.54}_{-0.46}$ \\
        &  \text{AD} = $-13.32^{+0.25}_{-0.18}$ & \text{AD} = $-13.59^{+0.70}_{-0.52}$\\
        \text { PG 1115+080 } & $\mathrm{A_1 A_2}$ = $+8.43^{+0.28}_{-0.29}$ & $\mathrm{A_1 A_2}$ = $8.45^{+0.53}_{-0.51}$ \\
        & $\mathrm{A_1 B}$ = $9.98^{+0.32}_{-0.27}$ & $\mathrm{A_1 B}$ = $10.21^{+0.76}_{-0.49}$ \\
        & $\mathrm{A_1 C}$ = $27.10^{+0.42}_{-0.42}$ & $\mathrm{A_1 C}$ = $27.07^{+0.69}_{-1.23}$ \\
        \text { WFI 2033-4723 } & $\mathrm{A_1 A_2}$ = $-0.97^{+0.26}_{-0.23}$ &  $\mathrm{A_1 A_2}$ = $-1.15^{+0.80}_{-0.73}$ \\
        &  $\mathrm{A_1 B}$ = $36.50^{+0.43}_{-0.32}$ & $\mathrm{A_1 B}$ = $36.31^{+0.98}_{-0.74}$ \\
        & $\mathrm{A_1 C}$ = $-23.15^{+0.27}_{-0.51}$ & $\mathrm{A_1 C}$ = $-23.17^{+0.65}_{-0.84}$ \\
        \text{SDSS 1206+4332} & \text{AB} = $111.93^{+1.14}_{-0.82}$ & \text{AB} = $112.39^{+2.29}_{-1.09}$ \\
        \hline
    \end{tabular}
    \label{tab:results}
\end{table*}

%%%%%%%%%%%%%%%%%%%%%%%%%%%%%%%%%%%%%%%%%%%%%%%%%%

%%%%%%%%%%%%%%%%%%%%%%%%%%%%%%%%%%%%%%%%%%%%%%%%%%

%%%%%%%%%%%%%%%%%%%%%%%%%%%%%%%%%%%%%%%%%%%%%%%%%%

%%%%%%%%%%%%%%%%%%%%%%%%%%%%%%%%%%%%%%%%%%%%%%%%%%

%%%%%%%%%%%%%%%%%%%%%%%%%%%%%%%%%%%%%%%%%%%%%%%%%%

% Don't change these lines
\bsp	% typesetting comment
\label{lastpage}
\end{document}